\documentclass{aa}
\usepackage{graphicx}
\usepackage{tabularx}
\usepackage{rotating}
\usepackage{txfonts}
\usepackage{comment}
\usepackage{lscape}
\usepackage{threeparttable}
\usepackage{xcolor}
\usepackage{subfigure}
\usepackage{caption}
\usepackage{float}
\usepackage{ulem}

\newcommand{\hii}{H\,\textsc{i}}
\newcommand{\hi}{H\,\textsc{i}\,21cm}
\newcommand{\apx}{$\sim$}
\newcommand{\pc}{$\%$ }

\newcommand{\eg}[1]{\citep[e.g.][]{#1}}
\newcommand{\kmps}{km~s$^{-1}$}
\newcommand{\p}[1]{$^{-#1}$}
\newcommand{\pp}[1]{$^{#1}$}

\newcommand{\halpha}{H$\alpha$}

\newcommand{\beq}{\begin{equation}}
\newcommand{\eeq}{\end{equation}}

\newcommand{\tspin}{{T$_{\rm spin}$}}

\newcommand{\Mstar}{$\textrm{M}_*$}
\newcommand{\Msun}{M$_\odot$}

\newcommand{\mjypb}{mJy~beam$^{-1}$}

\newcommand{\oiii}{[O III]$\lambda\lambda$4958,5007$\rm \AA$}
\newcommand{\sii}{[S II]$\lambda\lambda$6717,30$\rm \AA$}
\newcommand{\nii}{H$\alpha$ and [N II]$\lambda\lambda$6548,84$\rm \AA$}
\newcommand{\oii}{[O II]$\lambda\lambda$3726,28$\rm \AA$}

\begin{document}

   \title{The \hii\ absorption zoo: JVLA extension to $z \sim 0.4$}

   \author{Suma Murthy\inst{1,2},
          Raffaella Morganti\inst{2,1},
          Tom Oosterloo\inst{2,1},
          Filippo M. Maccagni\inst{3} 
         }

   \institute{Kapteyn Astronomical Institute, University of Groningen, P.O. Box 800,
9700 AV Groningen, The Netherlands \\
             \email{murthy@astro.rug.nl}
         \and
             ASTRON, the Netherlands Institute for Radio Astronomy, Oude Hoogeveensedijk 4, 7991 PD Dwingeloo, The Netherlands.
        \and 
    INAF - Osservatorio Astronomico di Cagliari, via della Scienza 5, 09047, Selargius, CA, Italy}

   \date{Received 17 June 2021, accepted 23 July 2021}


 \abstract{We present an \hi\ absorption study of a sample of 26 radio-loud active galactic nuclei (AGN) at $0.25 < z < 0.4$ carried out with the Karl G. Jansky Very Large Array. Our aim was to study the rate of incidence of \hii\ in different classes of radio AGN, the morphology and kinematics of the detected \hii, and the nature of the interaction between the \hii\ and the radio source at these redshifts. Our sample consists of 14 sources with sizes of up to tens of kpc and 12 compact sources ($<$ a few kpc) in the radio-power range 10\pp{25.7}~W~Hz\p{1} -- 10\pp{26.5}~W~Hz\p{1}. We detect \hii\ in five  sources, corresponding to a detection rate of \apx 19\%. Within the error bars, this agrees with the detection rate found at lower redshifts. We find that the rest-frame UV luminosities of most of the sources in the sample, including all the detections, are below the proposed threshold above which the \hii\ is supposed to have been ionised. An analysis of the optical emission-line spectra of the sources shows that despite their high radio powers, about one-third of the sample, including two detections, are low-ionisation sources. The radio continuum emission from the sources detected in \hii\ is unresolved at \apx 5 to 10~kpc scales in our observations, but shows extended structure on parsec scales. We  analysed the \hi\ absorption spectra of the detections to understand the morphology and kinematics of \hii. The absorption profiles are mostly complex with widths between the nulls ranging from \apx 60 \kmps\ to 700 \kmps. These detections also exhibit remarkably high \hii\ column densities in the range \apx 10\pp{21} cm\p{2} to 10\pp{22} cm\p{2} for \tspin$=$100 K and unit covering factor. Our modelling of the \hi\ absorption profiles suggests that in two sources the gas appears to be disturbed, and in three cases, including one with disturbed \hii, the majority of the absorption is consistent with it arising from an \hii\ disc. Despite the high radio power of our sources, we do not detect fast outflows. However, the optical emission lines in these detections show the presence of significantly disturbed gas in the nuclear regions in the form of very wide and highly blueshifted emission-line components. Since some of our detections are also low-ionisation AGN, it is quite possible that this disturbance is caused by the radio jets. Overall, our findings point towards a continuation of the low-$z$ trends in the \hii\ detection rate and the incidence of \hii\ in radio AGN up to $z \sim 0.4$. 
}


\titlerunning{\hii\ absorption to $z \sim 0.4$}
\authorrunning{Murthy et al.}

\maketitle

\section{Introduction} \label{introduction}

Understanding the distribution and properties of gas, in particular the cold neutral atomic and molecular phases, is crucial to understanding the evolution of galaxies. Molecular gas is now increasingly being detected in galaxies across a wide redshift range \citep[e.g.][]{Tacconi18,Neeleman18, Kanekar18, Decarli19,Aravena19,Neeleman20, Kanekar20}. Neutral atomic hydrogen has also been studied extensively via the 21cm emission line in the nearby universe. At higher redshifts, stacking experiments have provided information about the average \hii\ content in galaxies even out to $z$ \apx 1. \citep{Bera19, Chowdhury20b, Chowdhury21}. However, beyond the Local Universe, \hi\ emission has been detected from individual galaxies up to $z=0.376$ \citep{Fernandez16} with most of the studies limited to $z \lesssim 0.2$ \eg{Catinella08, Catinella15, Verheijen07, Hess19, Bluebird20, Gogate20}, although only with very deep integrations.

The detection of the \hi\ absorption line on the other hand depends on the strength of the background radio continuum and hence, can be achieved with short integration times, even at much higher spatial resolution, and higher redshifts where direct detection of emission is not yet   possible. Thus, \hi\ absorption has been used extensively to study atomic gas in quiescent \eg{Kanekar09, Gupta09, Kanekar14} and radio-loud active galaxies \citep[see][for a review]{Morganti18}.

In the case of radio-loud active galactic nuclei (AGN), \hi\ absorption enables us to trace cold gas close to the centre of the galaxy that is being affected directly by the AGN. At low redshifts, there have been many \hi\ absorption studies of several types of radio AGN \citep[e.g.][]{vanGorkom89,Morganti01, Vermeulen03, Gupta06, Curran08, Aditya18b}. \citet{Gereb15} and \citet{Maccagni17} did a detailed study of one such large sample of radio AGN (\apx 250 sources) observed with the Westerbork Synthesis Radio Telescope (WSRT). The radio power of the sources ranged from 10\pp{22.5} W Hz\p{1} to 10\pp{26.2} W Hz\p{1}. The study consisted of compact and extended radio sources in the redshift range $0.02 < z < 0.25$ and found \hii\ in a variety of morphologies, essentially an \hii\ `zoo'.

This study, with a 3$\sigma$ optical depth sensitivity of \apx 1\%, found a detection rate of \apx27\% across the entire range of radio powers covered. They found that radio sources at the low-power end ($L\rm_{1.4GHz}<10^{23}$ W Hz\p{1}) mostly show narrow absorption features indicative of gas settled in rotating discs. Broad blueshifted absorption features, which may represent unsettled gas, were detected only in higher power sources. They found that the sources with disturbed gas were also of compact radio
morphology (kiloparsec scale)  and rich in heated dust (as ascertained from their mid-infrared colours). Furthermore, they found that in compact radio sources the detection rate was \apx 30\%, while in the extended sources it was \apx 15\%. Similarly, dust-poor sources had a lower detection rate compared to the sources rich in heated dust. This study supported the findings of earlier work \eg{Vermeulen03, Gupta06} based on much smaller samples. A study of ionised gas in this sample further supported the finding that compact, high radio-luminosity sources cause more disturbance in the surrounding gas \citep{Santoro_thesis}. 

These results point to an interesting scenario where extended (older) radio sources no longer directly interact with the gas in the central region of the galaxy, while compact radio sources (young or restarted), which are still embedded within their host galaxies, strongly couple with the interstellar medium (ISM) and thereby induce a strong disturbance in the gas (negative feedback). This coupling is predicted by the simulations of radio jets expanding into the ambient ISM \eg{Wagner11, Mukherjee16, Mukherjee18a, Mukherjee18b, Bicknell18}. This suggests that a radio AGN may have a significant impact on the host galaxy over a certain period of its life span and thus contribute to this negative feedback effect. Hence, it is desirable to conduct studies such as that of \citet{Maccagni17} over a range of redshifts, which would tell us how the effect of radio jets on the ISM varies as a function of cosmic time and of the type of the radio source.

However, as we go to higher redshifts, the number of radio sources detected in \hi\ absorption decreases. One of the main reasons proposed for this is the selection effect, which causes radio sources studied at higher redshift to also have high restframe UV luminosity, which in turn ionises the cold gas \eg{Curran08}. Another reason could be the intrinsic decrease in the cold gas content at higher redshifts \eg{Aditya16, Aditya18b}. Blind \hi\ absorption surveys at high redshifts can address the issue of selection effects. Such surveys have started and have already found cold gas reservoirs in low-luminosity radio sources even at $z > 1$ \citep{Chowdhury20b} and suggest that selection effects could indeed be the cause of the apparent small number of the detections at high redshifts. At the moment, most studies of \hi\ absorption at higher redshifts have focused on compact radio sources or flat-spectrum sources for which the detection rate at low redshifts has been found to be higher. Thus, we lack a large sample of radio sources of different radio powers and morphologies corresponding to different stages of evolution of the radio sources at these redshifts that would allow us to make an analysis similar to that in  \citet{Maccagni17}. 

This work is a continuation of the study by \citet{Gereb15} and \citet{Maccagni17} to higher redshifts. In this paper we present a pilot study of a sample of 26 radio sources selected using similar criteria. We  used the Karl G. Jansky Very Large Array (JVLA) covering down to frequencies $\sim 1$ GHz, corresponding to $z \sim 0.4$ for the \hi\ line, to search for \hii\ in these radio sources.

We present the sample selection, observations and data reduction in Sect. \ref{sec:observations} and describe the properties of the sample at different wavelengths in Sect. \ref{sec:sample}. We present the \hii\ and continuum results in Sect. \ref{sec:results}. Finally, we discuss the results in Sect. \ref{sec:discussion}, where we also compare our findings with other studies from the literature and summarise in Sect. \ref{sec:summary}.

We have assumed a flat universe with H$_0$ = 67.3 km s$^{-1}$Mpc$^{-1}$, $\Omega_{\Lambda}$ = 0.685, and $\Omega_M$ = 0.315 \citep{Planck14} for all our calculations. For the redshift range $0.26 < z < 0.4$ covered by our sample, 1$''$ ranges between 4.2 kpc and 5.5 kpc. 

\section{Sample selection, observations, and data reduction}\label{data}
\label{sec:observations}

Our sample was constructed by cross-matching the  Faint Images of the Radio Sky at Twenty-Centimeters (FIRST; \citealt{Becker95}) and  the Sloan Digital Sky Survey Data Release 13 (SDSS DR13; \citealt{Albareti17}) catalogues, similar to those used by  \citet{Gereb15} and \citet{Maccagni17} in the redshift range 0.26 -- 0.4, the highest redshift accessible with the JVLA L band. We used the telescope in A configuration since this provides the highest possible spatial resolution, corresponding to a few kpc, at the redshifted \hii\ 21cm frequency. The observations were carried out under the projects 15A-065 (performed between 21 August and 20 September 2015), 18A-425 (performed on 17 May 2018) and 19A-031 (performed between 04 August and 18 September 2019).

We selected a total of 30 radio sources whose FIRST flux density ranged between 400 mJy beam\p{1} and 800 mJy beam\p{1}. The flux cutoff was chosen to limit the integration time to \apx 30 minutes per source without compromising on the optical depth sensitivity (\apx 1\pc without any spectral smoothing, comparable to \citealt{Maccagni17}).

We observed suitable flux calibrators at the start of each observing block. Each target was observed for 30 minutes. These target observations were interleaved with observations of the corresponding phase calibrator. We used a 32 MHz band subdivided into 1280 channels, giving us an unsmoothed spectral resolution of around 6.5 \kmps\ to 8~\kmps, depending on the redshift. We used the second WIDAR IF to cover the radio continuum in the first observing run (15A-065), while in the later runs we only used the 32 MHz band for spectral line observations.

We reduced the data using the  classic Astronomical Image Processing
Software \citep[AIPS;][]{Greisen03}. After flagging the bad data through visual inspection, we used the flux and phase calibrators to determine the antenna-dependent gain and bandpass solutions. Then we used self-calibration to improve the gain solutions. Here, we performed a few cycles of imaging and phase-only self-calibration and then one round of amplitude and phase self-calibration and imaging. We subtracted the continuum model from the calibrated visibilities and flagged the residual data for  radio frequency interference (RFI). Then, to subtract any residual continuum emission, we fit a second-order polynomial to the line-free channels of each visibility spectrum. We produced the final line cube after de-redshifting the $uv$ data to the systemic velocity of the target using the redshift provided by the SDSS DR13 catalogue.

The continuum and spectral beam sizes and RMS noise levels we achieved are listed in Table \ref{radio_obs}. We made the continuum images by averaging all the line-free channels and using two weighting schemes: robust weighting of --1 (to produce an image of the continuum structure at a high enough spatial resolution without compromising on sensitivity) and natural weighting (giving the maximum sensitivity). The spatial resolution of the images we obtain ranges from \apx 1$''$ to 3.8$''$ with ROBUST $-1$ weighting and between 2$''$ and 5$''$ for natural weighting. The range of beam sizes depends on the RFI conditions (i.e. how much flagging was needed) during the observations. 

The spectral cubes were made with natural weighting and with the same restoring beam as the naturally weighted continuum image.  We reach a typical continuum RMS noise of a few hundred $\rm \mu$Jy beam\p{1}. The typical RMS on the spectral cubes is \apx 2.5 mJy beam\p{1} per 6.5 to 8 \kmps\ channel.

The RFI at the observed frequencies has significantly affected the quality of the data. The data on three sources, SDSS\,J085451+621850 at 1120.7 MHz, SDSS\,111141+355337 at 1088.6 MHz, and SDSS\,J114539+442022 at 1092.8 MHz, are entirely affected by RFI, and hence are unusable. For many other targets, large chunks of channels are affected by RFI. However, we  included them in our analysis since if the \hi\ absorption profile is a few hundred  \kmps\ wide (as is often  the case for \hii\ in radio galaxies) the line, if present, would be expected to be detected in the uncontaminated channels around the contaminated ones. However, we  note that very narrow ($\ll50$ \kmps) absorption features may still be lost. Finally, for SDSS\,J112434+161651, the SDSS redshift  was found to be incorrect (see also Sect. \ref{sec:OpticalProperties}). Thus, in total our final sample consists of 26 sources. For two more sources (SDSS\,J075622+355442 and SDSS\,172109+354216), the RFI situation is quite bad and the spectra are barely acceptable. In the case where we reject them, the sample consists of 24 objects.

For calculating optical depths and the \hii\ column density (detection and limits) we use the naturally weighted continuum images and cubes. For the discussion of the  morphology of the radio continuum emission we use the higher resolution continuum images made with ROBUST $-1$ weighting. The spectra of the \hi\ non-detections are presented in Appendix \ref{sec:HIundetected} and the continuum images are presented in Appendix \ref{sec:ContImages}. The observation details including the spatial resolution of the naturally weighted continuum images and those made with ROBUST $-1$ weighting are given in Table \ref{radio_obs}.

To complement the \hi\ search, we also analysed the optical spectra available from the SDSS DR13 database to probe ionised gas kinematics in the sources in our sample. The line-flux measurements provided by SDSS are all estimated by fitting a single-component Gaussian to the emission lines. However, in many cases, the lines are better fit with multiple Gaussian components. Hence, we re-analysed the spectra. We considered the \oiii\ doublet, \nii, the  \sii\ doublet, and the \oii\ doublet whenever available for our analysis. For the fitting of the lines we largely followed the approach described in \cite{Santoro18}. However, we did not fit and subtract a stellar continuum as in \citet{Santoro18}, but instead fit a second-order polynomial locally around the emission lines simultaneously along with the Gaussian components.

For the \oiii\ doublet we assumed a kinematic component consisting of two Gaussians separated by fixed widths corresponding to the width between the doublet lines, and of the same FWHM. The line ratio of the 5007$\rm \AA$ line to the 4959$\rm \AA$ line was fixed to 3.0 for each component. The fitting was done using the python implementation of the Markov chain Monte Carlo method by \citet{Foreman13}. For the rest of the lines, we kept the number of kinematic components the same as that obtained for \oiii\ doublet and varied all other parameters according to the emission line being fit. For the \nii\ lines we fixed the ratio of the line strength of [N II]$\lambda$6584$\rm \AA$ to [N II]$\lambda$6548$\rm \AA$ to 3:1; for the \oii\ doublet the ratio of the 3728$\rm \AA$ to the 3726$\rm \AA$ line was fixed to 1.5:1. We then measured the flux of the line by integrating the area under the Gaussian components. In the case of multiple components the measured flux is the total flux from all the components. The full width at half maximum (FWHM) values we measured for the kinematic components of ionised gas for the sources detected in \hi\ absorption are listed in Table \ref{optical_table_det}, and the flux measurements for all the sources are presented in Table \ref{optical_table}.\\

For SDSS\,J112434+161651 we derive a redshift \apx0.67 using the weak \oiii\ doublet and the \halpha\ lines, while the redshift listed in DR13 is \apx0.28. Thus, as mentioned earlier, we  excluded this source from our sample.

\section{Properties of the sample}
\label{sec:sample}

\subsection{Radio properties}
\label{sec:RadioProperties}

\begin{figure}
    \includegraphics[width=9cm]{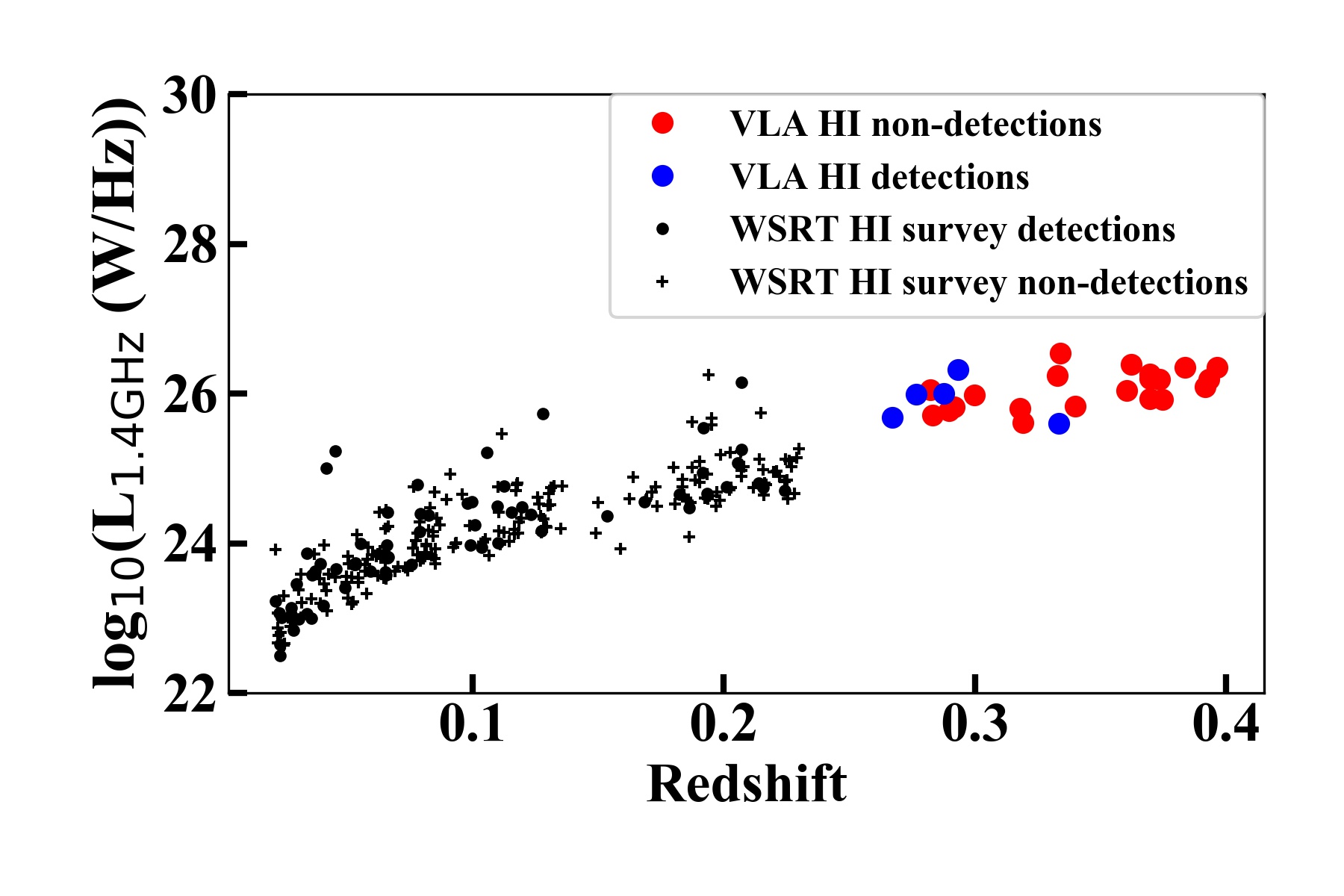}
    \caption{Comparison of the radio power of our sample with that of \citet{Maccagni17}.
    }
    \label{fig:radiolum}
\end{figure}

Figure \ref{fig:radiolum} shows the distribution of our sources in the radio luminosity-redshift space  compared to \citet{Maccagni17}. The rest-frame 1.4 GHz radio luminosity of the objects of our sample ranges from $4 \times 10^{25}$ W Hz\p{1} to $4 \times 10^{26}$ W Hz\p{1}, comparable to the most powerful sources in the sample of \citet{Gereb15} and \citet{Maccagni17}. Most of the studies at higher redshifts have radio powers ranging from $10^{27}$ W Hz\p{1} to $10^{29}$ W Hz\p{1} \eg{Aditya16, Aditya17, Aditya18a, Curran08, Curran11b, Curran17}, higher than that of the sources in our sample. Hence our sample serves as a bridge between the low- and high-redshift associated \hi\ absorption studies. Within our sample, we find that the detections are concentrated in the lower radio-power portion of our sample. 

The radio continuum images  of the sample are presented in Fig. \ref{fig:continuum_maps}. Of the 26 sources, 12 sources are unresolved in our JVLA observations.  These images show the radio continuum emission on spatial scales ranging from 5 kpc to 20 kpc. However, high spatial resolution very long baseline interferometry (VLBI) images with milliarcsecond resolution at higher frequencies are available at the Astrogeo database\footnote{\url{http://astrogeo.org/vlbi_images/}. The VLBI images shown in this paper were made by Leonid Petrov.}. Ten of these sources are unresolved with the JVLA, and one source remains unresolved even on VLBI scales (corresponding to \apx50 pc to 200 pc), while the rest exhibit extended features at scales of a few tens of mas. Of the sources unresolved in our sample that have a classification available in the literature, one is classified as a compact steep spectrum source (CSS), another  is classified as a gigahertz peaked-spectrum (GPS) source, and one other has been classified as a peaked-spectrum source \citep{Callingham17}. These types of sources are considered  young radio sources \citep[e.g.][]{Odea98}. The rest are classified  as flat-spectrum sources or blazars. The sources that are compact at VLBI scales may still have extended low surface-brightness emission on kiloparsec scales at lower frequencies. The presence of this emission can only be investigated with deep low-frequency observations which are planned using the LOFAR surveys. Such a morphology, of bright compact emission along with a diffuse extended low-frequency emission component, is usually associated with young or restarted sources where the kiloparsec-scale diffuse emission is not the dominant component of radio emission. Thus, we argue that even if this emission is present in these sources, they will still remain core-dominated. 

All our \hi\ detections are unresolved at the resolution of our images ($\lesssim 5$ kpc), although they exhibit more complicated structures on VLBI scales. A more detailed discussion will be presented in Sect. \ref{sec:discussion}.

\subsection{Optical properties}
\label{sec:OpticalProperties}

\subsubsection{Optical magnitudes}
\begin{figure}
    \includegraphics[width=9cm]{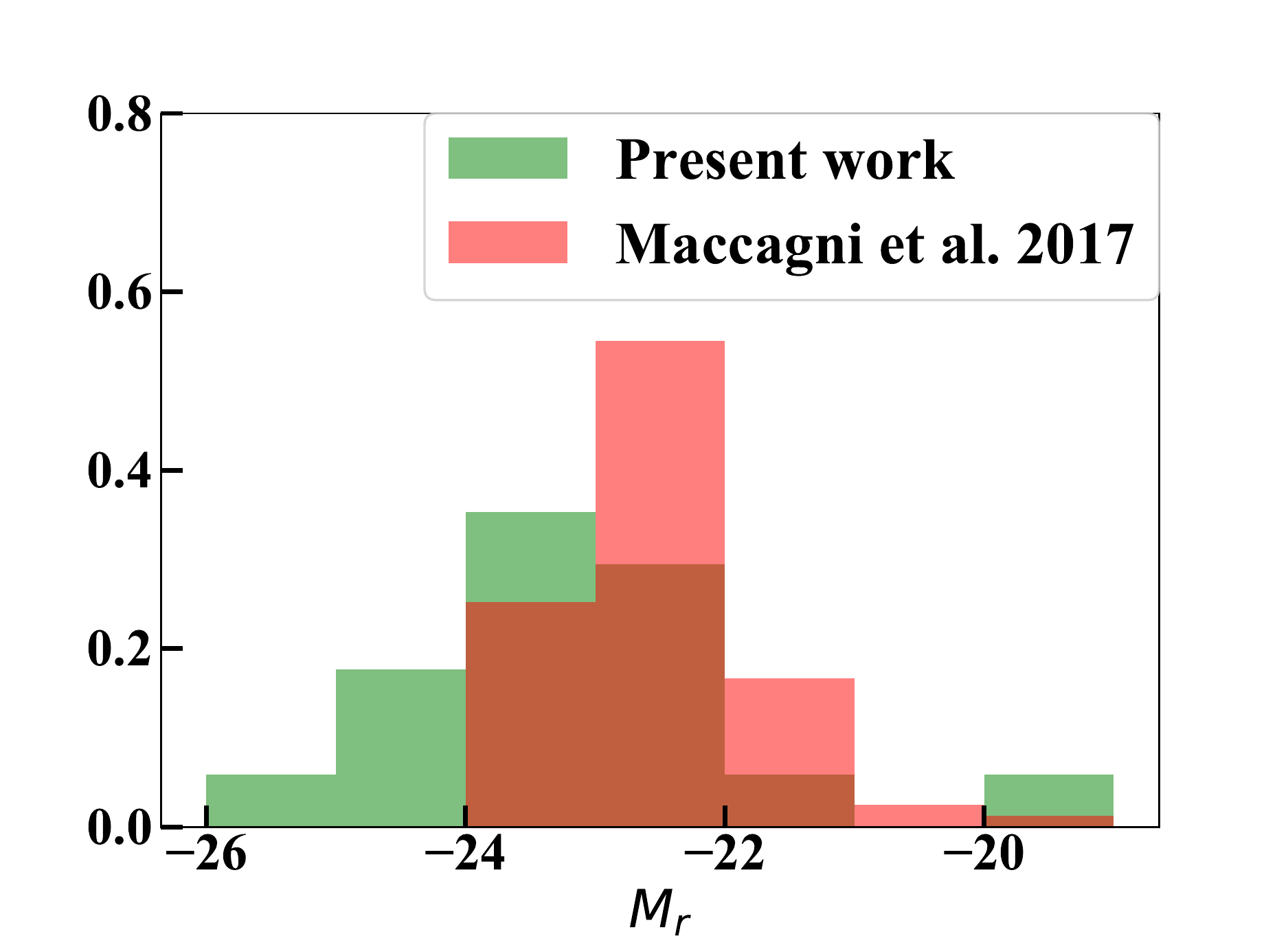}
    \caption{Normalised histogram showing a comparison of the K-corrected absolute r-band magnitudes of our sample with that of \citet{Maccagni17}.
    }
    \label{fig:optical_mag}
\end{figure}

Figure \ref{fig:optical_mag} shows the distribution of K-corrected \citep{Chilingarian10,Chilingarian12} r-band absolute magnitudes of the sources in our sample  compared to that in \citet{Maccagni17}. The magnitudes range from $-23.4 < M_{\rm r} < -19.6$. In comparison, the magnitudes of the sources studied by \citet{Maccagni17} lie in the range $-24 < M_{\rm r} < 18.5$. Thus, in terms of stellar mass, of which $ M_{\rm r}$ is a good proxy, our sample overlaps with that of \citet{Maccagni17}.

\subsubsection{Ionised gas and optical spectra}

Emission from the ionised gas present in the surroundings of the AGN provide additional useful information on the coupling of radio jets with the ISM. Many studies \eg{Santoro18} have shown that broad kinematic components, indicative of gas with disturbed kinematics, are seen more often in the emission lines of ionised gas than in \hi\ absorption profiles. Thus, they seem more suited to trace the jet-ISM interaction. However, the ionised gas outflows are typically less massive than the cold gas outflows. Hence it is useful to combine these two diagnostics (ionised gas and \hii) for a more complete picture of the condition of gas in the nuclear region of radio AGN. 

We find that for half of the sources in our sample, the fitting of the line profiles requires two components, suggesting a more complex kinematics of the ionised gas. We used the flux ratios derived using these measurements to construct the BPT diagram shown in Fig. \ref{fig:bpt_sample}. One of the \hi\ detections, SDSS\,J145845+372022, is a blazar with a featureless optical spectrum \citep{Massaro09}, and hence could not be shown in the BPT diagram. We find that despite being powerful radio sources, ten of the sources in the sample lie in the star-forming -- composite or LINER region of the diagram. This also includes two of the sources detected in \hi\ absorption. However, the equivalent width of the \oiii\ line for eight of these sources is greater than 5$\rm\AA$ indicating that these are high-excitation radio galaxies \citep{BestHeck12}. The \oiii\ lines are not detected for the other two sources.

\begin{table*}[]
\caption{The \oiii\ line widths  of the AGN detected in \hi\ absorption.}
\centering
\begin{tabular}{ccccc}\hline
Source & $z$ & FWHM$\rm_{[OIII]}$ & FWHM$\rm_{[OIII]}$ & $v\rm_{component 2}$ \\
 &  & component 1 & component 2 \\
 &  & (\kmps) & (\kmps) & (\kmps) \\
 (1) & (2) & (3) & (4) & (5)\\
\hline
J090151+030423 & 0.2875 &  390  $\pm$ 51   &  870  $\pm$  260  &  -650 $\pm$ 150 \\
J142904+025140 & 0.2931 &  423  $\pm$ 141  &  -                & - \\ 
J145239+062738 & 0.2671 &  705  $\pm$ 35   &  1410 $\pm$  117  &  -300 $\pm$ 160 \\
J145845+372022 & 0.3332 &  -               &  -                & - \\
J234107+001833 & 0.2767 &  1457 $\pm$ 94   &  -                & - \\
\hline
\end{tabular}    
\begin{tablenotes}
\item The columns are: (1) Source name, (2) SDSS DR13 redshift, (3) FWHM in \kmps\ of the narrow kinematic component which is at the systemic velocity of the AGN, (4) FWHM in \kmps\ of the second, wide component that is offset from the systemic velocity, (5) Centroid of the second component in \kmps.
\end{tablenotes}
\label{optical_table_det}
\end{table*}

\subsection{UV Luminosities}
\label{sec:UVluminosity}

We estimated the rest-frame UV luminosity of the sources following the prescription of \citet{Curran08}. We used the SDSS u and g bands to estimate the rest-frame 1216$\AA$ flux density ($S_{\rm 1216\AA}$) by fitting a power-law spectrum, and then estimated the luminosity by using the expression $4\pi D_L^2S_{\rm 1216\AA}/(1+z)$. For some of the sources, far-UV and near-UV magnitudes are available from the Galaxy Evolution Explorer (GALEX). The UV luminosities derived using GALEX measurements are comparable to those obtained from the SDSS magnitudes. However, we note that using the SDSS magnitudes for the purpose would ensure that the estimated values correspond to the emission from the nuclear region, while the GALEX measurements may include emission from the entire galaxy. Hence we  used the UV luminosity values estimated from the SDSS magnitudes. The UV luminosity of our sources as a function of redshift is shown in Fig. \ref{fig:uvlum}. The UV luminosity of our sources ranges from $10^{18}$ W Hz\p{1} to $10^{23.2}$ W Hz\p{1}. All the sources with the exception of one have UV luminosities below $10^{23}$ W Hz\p{1}, the cutoff proposed by \citet{Curran08} corresponding to the luminosity of an unobscured quasar. Other studies at similar redshifts   also consider sources with similar UV luminosities \eg{Aditya18b, Aditya18a, Aditya17}. At higher redshifts, however, the UV luminosities of most of the sources studied except, for example, \citet{Chowdhury20a} are above this threshold value.

\subsection{WISE colours}
\label{sec:wisecolours}

Figure \ref{fig:wise} shows the distribution of our sources in the Wide-field Infrared Survey Explorer (WISE) colour-colour plot. This plot can be used to classify the galaxies according to their dust content based on the colours derived from the three WISE bands (W1, W2, and W3). Colour W2$-$W3, corresponding to 4.6$\mu$m $-$ 12$\mu$m, is sensitive to the presence of dust;  W1$-$W2, which corresponds to 3.6$\mu$m $-$ 4.6$\mu$m, is sensitive to the presence of heated dust. We would expect an enhancement in both these colours in dust-rich galaxies hosting an AGN, while the dust-poor galaxies would not show any enhancement in either colour. This kind of plot is used by \citet{Maccagni17} to investigate the relation between the dust content of the host galaxy and the detection rate of \hii\ in absorption. They have found that the \hi\ detection rate is higher (\apx40\%) for sources in the mid-IR-bright region (i.e. the AGN with enhanced dust). \\

Given their high radio power, our sources are mostly located in the region corresponding to the sources dominated by AGN radiation, as can be seen in Fig. \ref{fig:wise}. All our \hi\ detections also fall in the same region of the plot where the detection rate was found to be higher by \citet{Maccagni17}.

\begin{figure}
    \includegraphics[width=9cm]{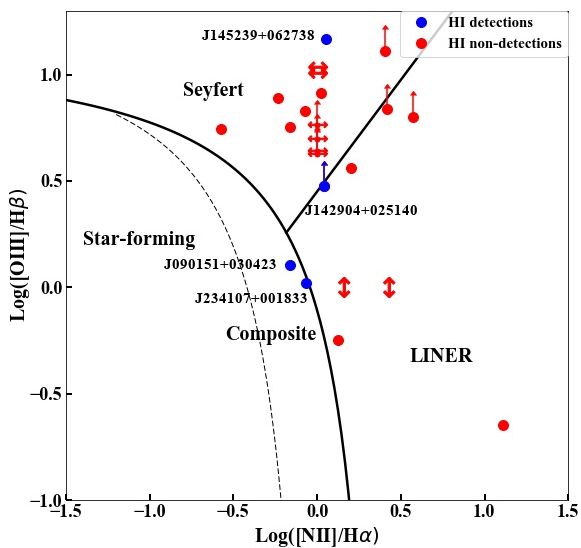}
    \caption{BPT diagram for the sample. Shown are the \hi\ detections (blue dots) and  the \hi\ non-detections (red dots). One of the detections, J1458+3720, does not have any ionisation lines, and hence could not be included in the diagram. Most of the sources that require double Gaussian components to fit their emission lines are also the more powerful AGN (in the Seyfert region of the diagram).}
    \label{fig:bpt_sample}
\end{figure}

\section{Results}\label{sec:results}

We detect \hi\ absorption in 5 out of the observed 26 sources. Below we present the statistics of the detection rate and  the multiwavelength properties of the detected objects. A comparison with other low- and high-redshift \hi\ absorption studies is done  in Sect. \ref{sec:discussion}. A more detailed modelling of some of the \hi\ absorption spectra is  done in Sect. \ref{sec:modelling}.

\subsection{\hi\ absorption detections}

We detect \hi\ absorption in 5 out of 26 sources for which a usable spectrum was obtained. This gives a detection rate\footnote{We estimate 1$\sigma$ error bars based on binomial statistics using \citet{Cameron11}} of 19\pc$^{+10\%}_{-5\%}$.

\begin{figure*}
    \includegraphics[width=8cm,height=5cm]{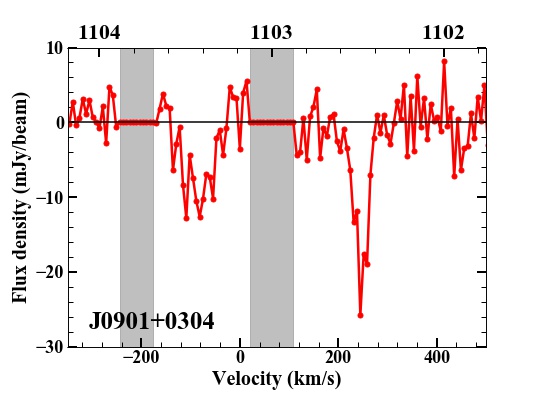}
    \includegraphics[width=8cm,height=5cm]{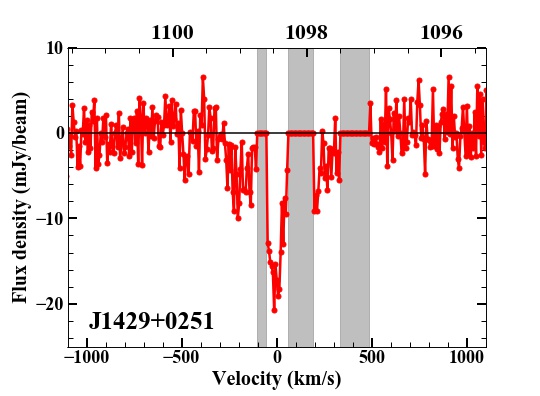}
    \includegraphics[width=8cm,height=5cm]{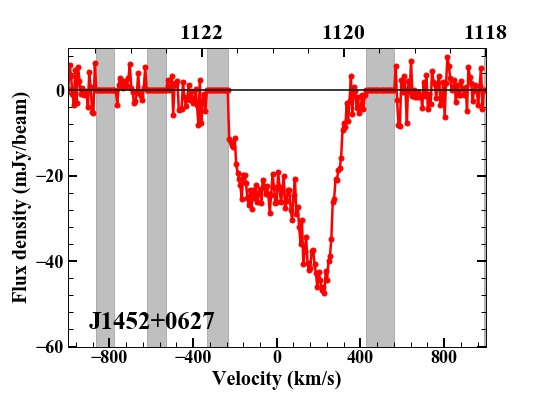}
    \includegraphics[width=8cm,height=5cm]{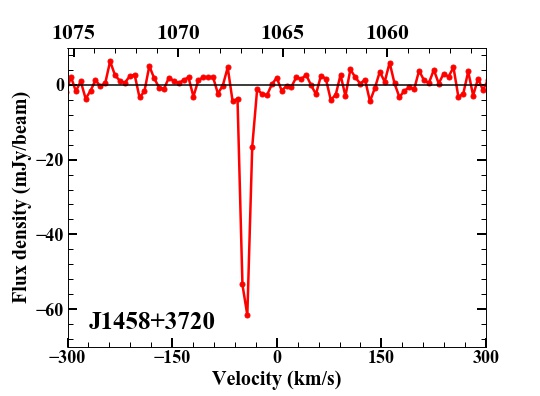}
    \centering\includegraphics[width=8cm,height=5cm]{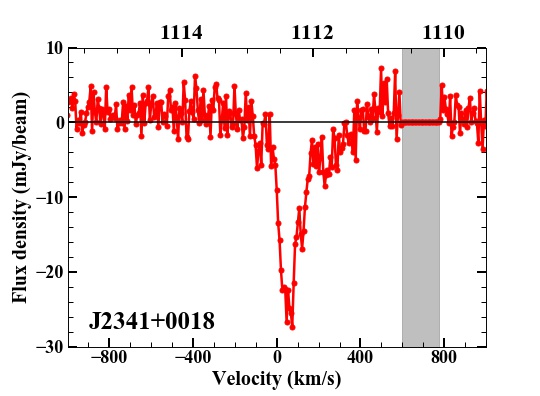}
    \caption{\hi\ absorption spectra of the sources. The heliocentric frequencies, in MHz, are given at the top of each panel. The frequency ranges affected by RFI are shown in grey.}
    \label{fig:det_spectra}
\end{figure*}

The \hi\ absorption spectra of the detections are shown in Fig. \ref{fig:det_spectra}. The absorption profiles we detect have a full width at zero intensity (FWZI) ranging from \apx60 \kmps\ to  \apx700 \kmps. The \hii\ column densities range between 8 $\times$ 10\pp{20} cm\p{2} and 10\pp{22} cm\p{2} assuming a \tspin\ of 100 K. The details of our observations are listed in Tables \ref{radio_table_det} and \ref{radio_table_nondet}. We   reached a 3$\sigma$ optical depth limit of $\lesssim$ 1\pc for most of the targets, similar to that achieved by studies at lower redshift \eg{Gereb15, Maccagni17}. In addition, this limit is almost constant over the redshift range covered. For the detections we   estimated the \hii\ column density  using the standard formula $N_{\hii} = 1.82 \times 10^{18} (T_{spin}/c_f) \int \tau dv$ cm$^{-2}$ where $\tau$ is the optical depth estimated by using the peak flux density of the radio continuum, \tspin\ is the gas spin temperature   assumed to be 100 K for all purposes, and $c_f$ is the covering factor assumed to be unity. For the non-detections, we  estimated the 3$\sigma$ upper limit to the \hii\ column density by first deriving a 3$\sigma$ optical depth limit using the peak flux density of the radio continuum and then assuming a Gaussian absorption profile  with an FWHM of 50 \kmps\ and a gas spin temperature of 100 K as has been done by \citet{Maccagni17}. Two of the \hii\ detections have already been reported as part of other studies (SDSS\,J90151+030423 and SDSS\,J145845+372022 by \citealt{Yan16} and \citealt{Aditya18b}, respectively)\footnote{These observations were carried out independently at almost the same time.}. We make a comparison with these studies in Sect. \ref{sec:detections}.

\begin{figure}
    \includegraphics[width=9cm]{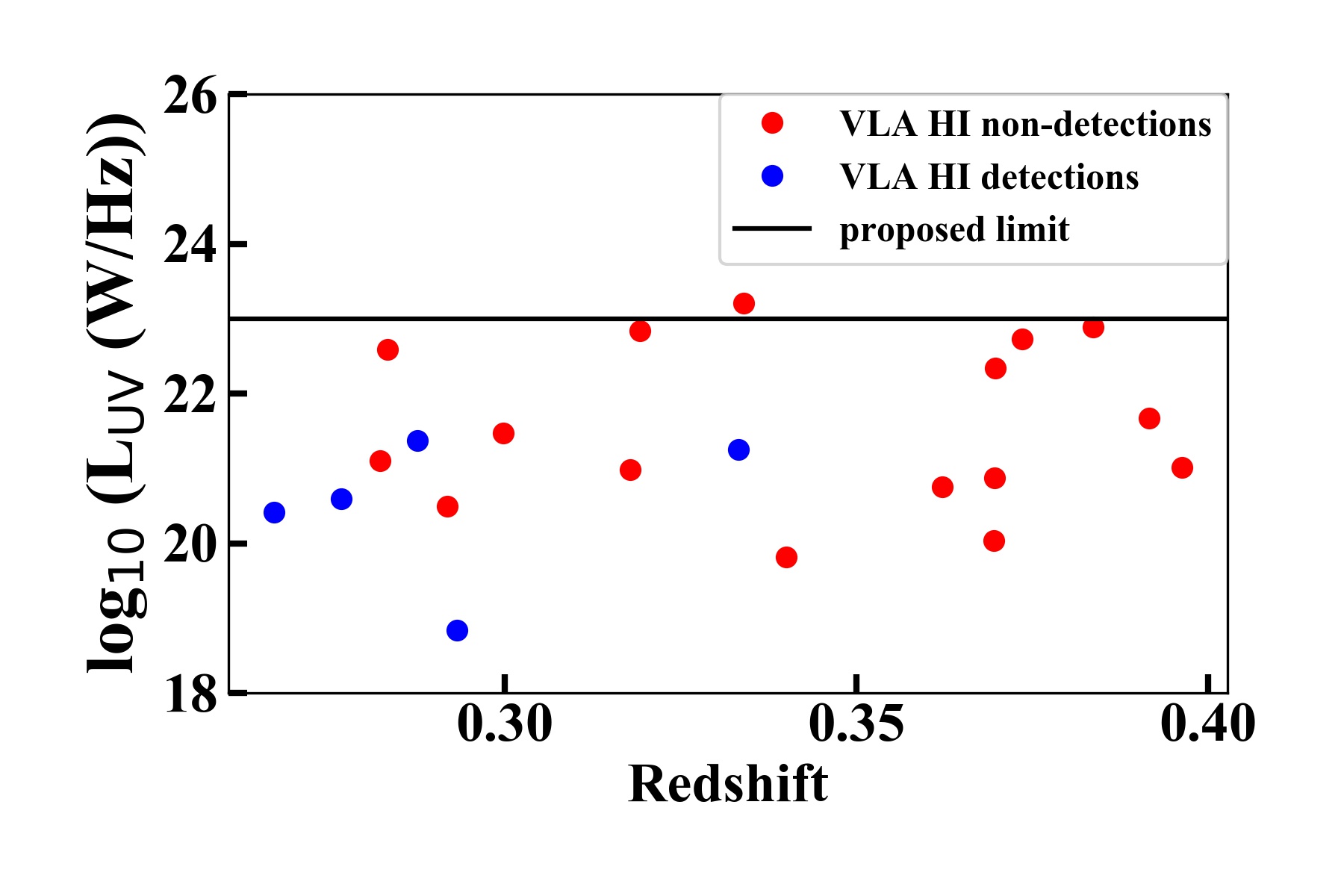}
    \caption{UV luminosities of the sources in the sample. See the text for details of the limit indicated by the horizontal line.}
    \label{fig:uvlum}
\end{figure}

\begin{figure}
    \includegraphics[width=9cm]{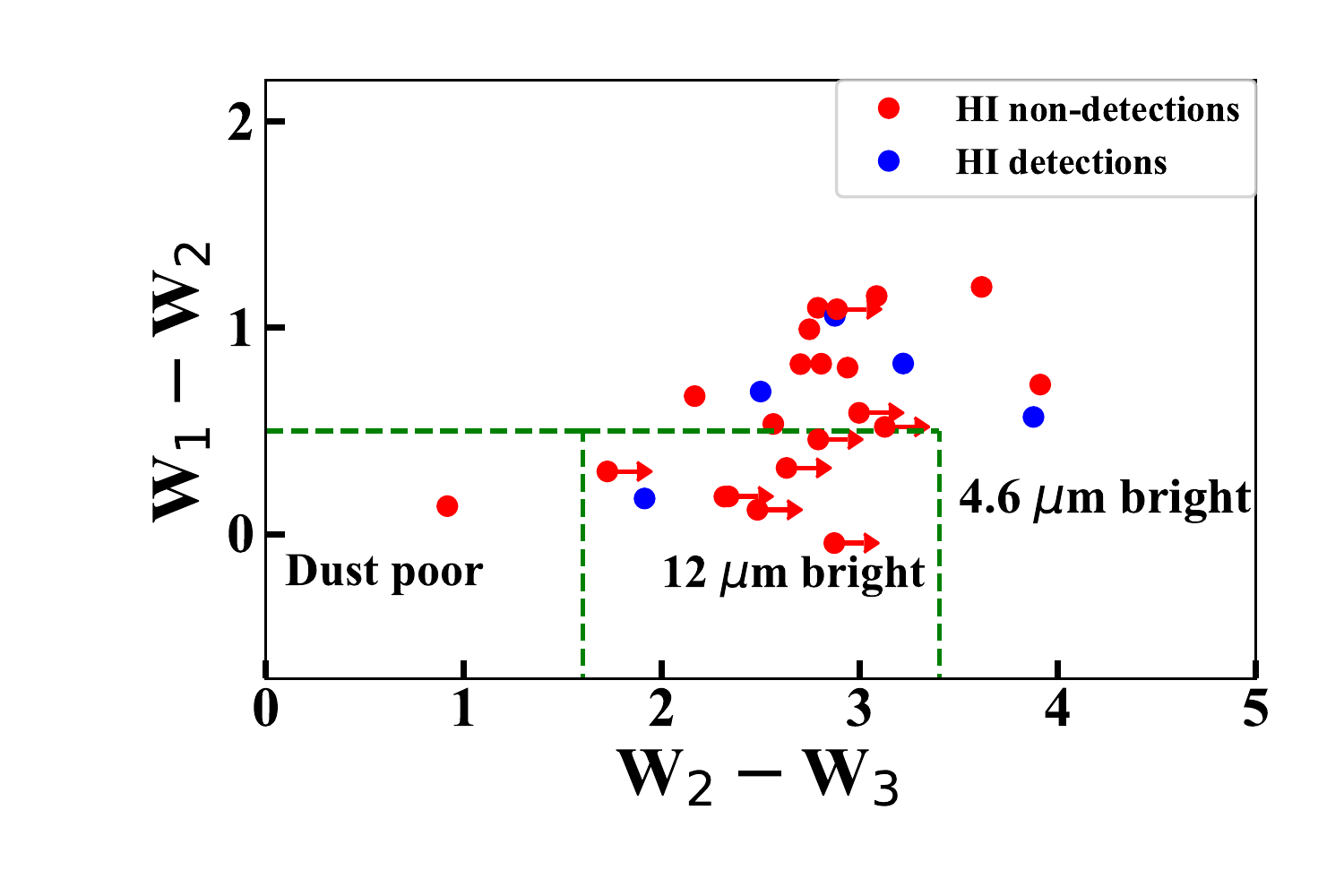}
    \caption{WISE colour-colour plot for our sample. The dashed lines indicate the cutoffs of the WISE colours we used to classify our sources.}
    \label{fig:wise}
\end{figure}

Figure \ref{fig:bpt_sample} shows where the \hi\ detections are located in the BPT diagram. These sources show a variety of optical properties, from composite or LINER to Seyfert-like (i.e. high-excitation) spectra. This is in agreement with that found for the sample presented in \citet{Maccagni17} where there was no significant difference in the \hi\ detection rate between Seyfert galaxies and LINERs \citep[see also][]{Santoro_thesis}. Furthermore, in Fig. \ref{fig:uvlum} we see that the rest-frame UV luminosity of the \hi\ detections is below the proposed threshold of 10\pp{23} W Hz\p{1}. Additionally, as can be seen in Fig. \ref{fig:wise}, in the WISE colour-colour diagram, the detections lie in the region corresponding to high detection rate at low redshifts.

\subsection{Properties of the HI detections}
\label{sec:detections}

\subsubsection{SDSS\,J090151+030423 (PKS~0859+032)}
\label{sec:0901_results}
Our JVLA \hi\ absorption spectrum of SDSS\,J090151+030423 (hereafter J0901+0304) is shown in Fig. \ref{fig:det_spectra}. The absorption has two components: a shallow component, blueshifted by \apx80 \kmps\ from the systemic velocity, and a narrow deep absorption \apx240 \kmps,\ redshifted with respect to the systemic velocity. Both the components have an FWZI of \apx120 \kmps. The \hi\ absorption was already reported from independent observations by \citet{Yan16}. Even though the frequency range is affected by RFI, multiple observations with the VLA and the GMRT \citep{Yan16}, and the JVLA (this work) agree on the two-component profile. Additionally, in the GMRT observations of \citet{Yan16}, where the band is unaffected by RFI, a shallow broad absorption component, \apx 140 \kmps\ wide, blueward of the narrow absorption is present. The quality of our data does not allow us to confirm this feature.

\citet{Yan12} have performed optical observations of this radio source and they report a slightly different redshift of $z=0.2872$ compared to that provided in the SDSS DR13 catalogue: $z = 0.2875$. From their optical and NIR observations, \cite{Yan12} further suggest that the host galaxy is an Sc galaxy \citep[based on the SED templates from][]{Mannucci01} and its optical and NIR images suggest that it may be an interacting system with some diffuse structure extending to the north-east of the main galaxy. We discuss the interpretation of the absorption profile also taking into account the difference in the systemic velocity in Sect. \ref{sec:0901_results}.

\begin{figure*}
\centering
   \includegraphics[width=7cm]{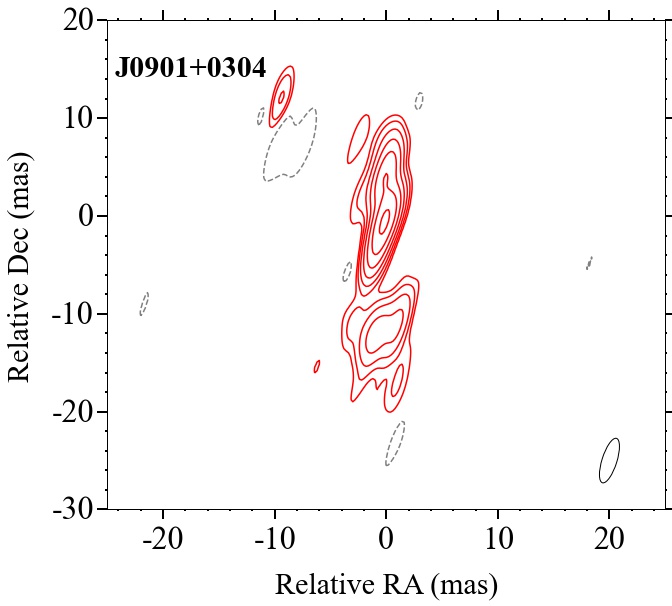}
   \includegraphics[width=7cm]{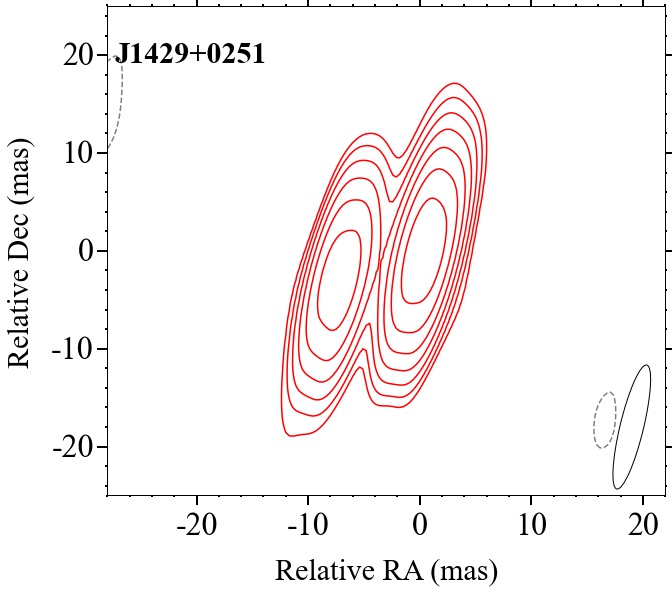}
   \includegraphics[width=7cm]{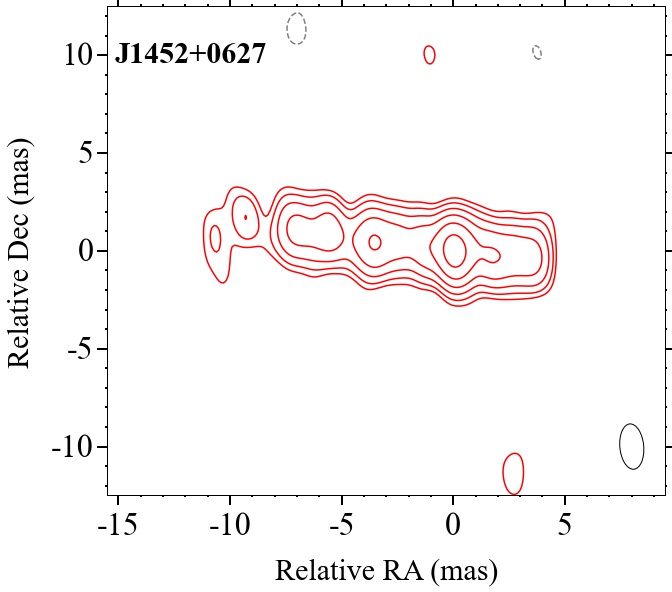}
   \includegraphics[width=7cm]{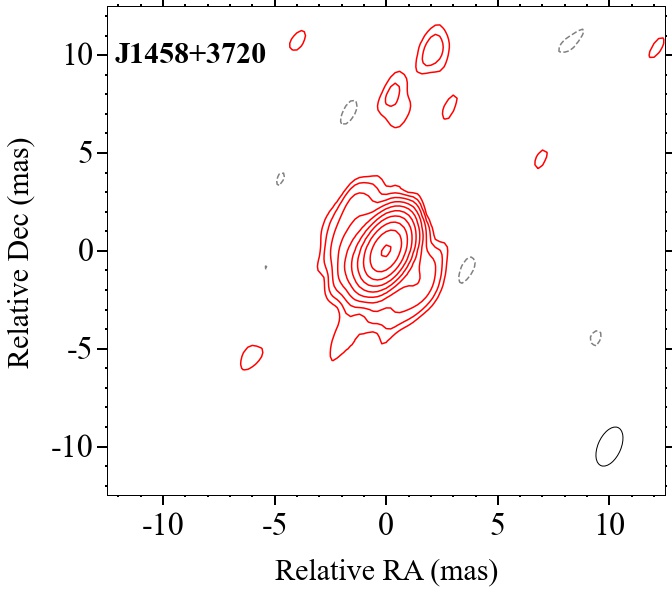}
   \includegraphics[width=7cm]{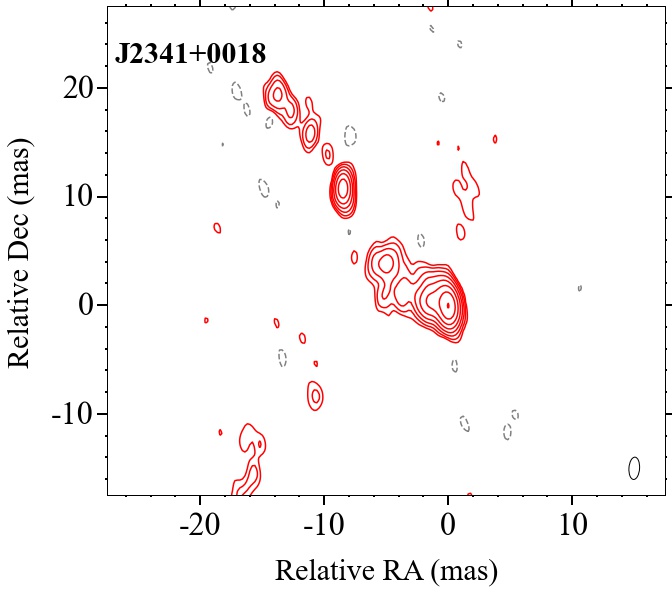}
    \caption{ VLBI continuum images of the detections at the highest resolution available, taken from the Astrogeo VLBI image database. The contours start at 3$\sigma$ and increase by steps of 2. The 3$\sigma$ negative contours are shown in grey. All the images are at X band except that of J1429+0251, which is at C band. These images are from \citet{Petrov21}, \citet{Schinzel17}, \citet{Gordon16}, and \citet{Beasley02}  and are used to model the absorption in Sect. \ref{sec:modelling}. The radio core is assumed to be at the centre of the image (corresponding to an offset of 0 mas  in RA and in Dec) for all sources except for J0901+0304 for which the assumed centre is to the south of the centre of the image.}
    \label{fig:radio_cont}
\end{figure*}

The radio continuum is unresolved at the arcsec resolution of our observations (Fig. \ref{fig:continuum_maps} and Table \ref{radio_obs}). We measure a total flux density of 380 $\pm$ 40 mJy. Our measurement is consistent with that of \citet{Yan16}. We measure a peak optical depth of 7\% and derive a total \hii\ column density (including both the absorption features) of $(8.3 \pm 0.5) \times 10^{20}$ cm\p{2}. The high-resolution VLBI image (\apx 3 mas resolution, Fig. \ref{fig:radio_cont}) shows that the radio continuum is extended with a jet-like structure \apx100~pc in size, in projection. The source is classified as a GPS source by \citet{Yan16} suggesting that it is a young radio AGN.

The optical emission lines were fit with a narrow and a broad component (see Table \ref{optical_table_det}). The broad component is significantly blueshifted by \apx 650 \kmps, suggesting the presence of kinematically disturbed gas in the circumnuclear region. The optical line ratios place J0901+0304 in the composite region between LINERs and star-forming galaxies in the BPT diagram (see Fig. \ref{fig:bpt_sample}) implying that the optical AGN is not very powerful. Multiple kinematic components in the optical spectrum could be a source of uncertainty in the redshift estimate which may affect the interpretation of the absorption lines (see Sect. \ref{sec:modelling}). 

\subsubsection{SDSS\,J142904+025142 (PKS\,1426+030)}

The \hi\ absorption spectrum of SDSS\,J142904+025142 (hereafter J1429+0251) is shown in Fig. \ref{fig:det_spectra}. The RFI affecting a big chunk of the channels right at the centre notwithstanding, the line appears to be very broad, with a maximum FWZI of \apx 700 \kmps. The deeper component of the absorption appears to be \apx 100 to 200 \kmps\ wide and centred at the systemic velocity. The spectrum also shows shallow wings on either side. Since parts of the spectrum are affected by RFI, it is not possible to characterise the absorption profile completely. Using only the channels not affected by RFI, we measure a peak optical depth of 3\% and derive a lower limit to the \hii\ column density of  $(10.2 \pm 0.6) \times 10^{20}$ cm\p{2}. 

J1429+0251 is unresolved in our JVLA observations, and we measure a peak continuum flux density of 810 mJy beam\p{1}. The high-resolution C-band VLBI image is shown in Fig. \ref{fig:radio_cont}. The radio source consists of two lobes and is \apx 100 pc in projection. It has been classified as a peaked-spectrum source by \citet{Callingham17} suggesting that this source is a candidate young radio galaxy. 

The optical spectrum of J1429+0251 consists of a weak [OIII] doublet and [NII]+\halpha\ lines that are well fit with a single kinematic component. We do not see any sign of disturbance in the nuclear regions in the optical emission lines. In Fig. \ref{fig:bpt_sample} we see that the source lies along the border between the Seyfert and LINER regions of the BPT diagram.

\subsubsection{SDSS\,J145239+062738 (TXS\,1450+066)}
\label{sec:J1452}

The \hi\ absorption spectrum of SDSS\,J145239+062738 (hereafter J1452+0627) is shown in Fig. \ref{fig:det_spectra}. The radio source is unresolved in our JVLA continuum image. It has been classified as a flat-spectrum source \citep{Healey07} and as a quasar \citep{Souchay12}. The Hubble Space Telescope (HST) optical image shows that the host galaxy has a dust lane aligned along the radio source\footnote{The HST image is available at the Mikulski Archive for Space Telescopes (MAST).}. The galaxy also exhibits a tidal tail suggesting that it may have undergone a merger.

The \hi\ absorption profile is significantly wide with a FWZI \apx800 \kmps. 
We also note that the profile is centred on the systemic velocity with the peak absorption redshifted by \apx200 \kmps. A more shallow absorption component is seen on the blueshifted side. This happens to be a high-opacity absorber with a peak optical depth of \apx 20\%. We derive a high total \hii\ column density of $(13.2 \pm 0.2) \times 10^{21}$ cm\p{2}, two orders of magnitude higher than the typical \hii\ seen in absorption in radio galaxies. This is, however, similar to the \hii\ column density measured in the nuclear region of galaxy mergers hosting radio AGN \citep{Dutta18}. 

The  high-resolution VLBI X-band image is shown in Fig. \ref{fig:radio_cont}. The radio continuum is about \apx50 pc in projection. The VLBI image shows multiple bright components. It could either be a core with the jets on either side seen almost in the plane of the sky, or a core and a one-sided jet. A detailed spectral index study clarifying the morphology is not available. 

 We find that J1452+0627 is a high-ionisation source and is located in the Seyfert region of the BPT diagram (Fig. \ref{fig:bpt_sample}). From the fitting of the optical emission lines from the SDSS spectrum, we find that the lines, except the \oii\ doublet, consist of two kinematic components (see Table \ref{optical_table}). Both the components are very wide (\apx700 \kmps\ and \apx1400 \kmps). The second component is blueshifted by \apx300 \kmps. These results indicate significant disturbance in the nuclear region. The width of the narrower kinematic component is close to the width of the \hi\ absorption line. This suggests that it might   be possible that both the components of gas (ionised and neutral) are being affected by the radio jets, as has been seen in other cases at low redshifts \citep[e.g.][see also Sect. \ref{sec:modelling}]{Morganti09c}.

\subsubsection{SDSS\,J145845+372022 (B3\,1456+375)}

The \hi\ absorption spectrum of SDSS\,J145845+372022 (hereafter J1458+3720) is shown in Fig. \ref{fig:det_spectra}. The absorption profile is very narrow with FWZI \apx 60 \kmps\ and is blueshifted by \apx 40 \kmps. We measure a peak optical depth of 33\% and an integrated \hii\ column density of $1.2 \pm 0.1) \times 10^{21}$ cm\p{2}. This absorption has been reported by \citet{Aditya18b}. They too find the absorption to be blueshifted. Our continuum flux density measurement agrees with that reported by \citet{Aditya18b} within the errors. The total \hii\ column density we measure after a reanalysis of their data (proposal ID: ddtB190) is consistent with our JVLA estimate given above, although the value reported in \citet{Aditya18b} is slightly lower.

J1458+3720 is classified as a blazar \eg{Massaro09, Dabrusco14}. Consistent with the classification as a blazar, the optical spectrum and, in particular, the continuum looks featureless, possibly due to the power-law emission from the nuclear region dominating over the stellar component. The redshift has been estimated from weak absorption lines which may have contributed to the uncertainty in the redshift. At VLBI scales, the radio source shows two extended features in addition to a bright core. The JVLA continuum image is shown in Fig. \ref{fig:continuum_maps} and the high-resolution VLBI image is shown in Fig. \ref{fig:radio_cont}.

\subsubsection{SDSS\,J234107+001833 (PKS\,2338+000)}

The \hi\ absorption spectrum of SDSS\,J234107+001833 (hereafter J2341+0018) is shown in Fig. \ref{fig:det_spectra}. The profile is broad with an FWZI \apx 550 \kmps. We find that the peak absorption is redshifted from the centre by \apx 50 \kmps. The absorption profile is complex, consisting of relatively shallow red- and blueshifted components along with the deep absorption.  We obtain a peak optical depth of 6.4\% and a total \hii\ column density of (1.8 $\pm$ 0.6) $\times$ 10\pp{21} cm\p{2}.

The radio source is unresolved in our observations and is smaller than \apx10 kpc. The high-resolution VLBI X-band image is shown in Fig. \ref{fig:radio_cont}. The projected size is only \apx 120 pc as measured in the X-band image. We used the VLBI X- and S-band images to make a spectral index map (not shown) and found that the radio continuum has an inverted-spectrum core and a one-sided jet. We measure a flux density of 443 \mjypb\ with the JVLA at 1.11 GHz. If we are to extrapolate the measured flux densities of the inverted-spectrum core (\apx 70 mJy) and the steep spectrum jet (\apx 270 mJy) at X band to the redshifted \hi\  frequency of 1.11 GHz by assuming spectral indices of 0.3 and $-$0.7 respectively (spectral indices measured between X and S bands), we would expect to measure a flux density of \apx500 mJy, which is in agreement with our measurement at 1.11 GHz. Hence, we conclude that the dominant component of the radio emission is concentrated in the central \apx120 pc. 

The SDSS spectrum of J2341+0018 shows strong emission lines on top of stellar continuum. Therefore, it is unlikely to be a blazar source, as has been suggested by \citet{Dabrusco14}. We find that all  four emission lines (the \oiii\ doublet, \nii\ lines, the \sii\ doublet, and the \oii\ doublet) are well fit with a single kinematic component. However, these lines are significantly broad (FWHM \apx 1400 \kmps), indicating disturbed kinematics. Furthermore, as can be seen from the location of the galaxy in the BPT diagram (Fig. \ref{fig:bpt_sample}), we find that optically it is a low-ionisation source and lies in the composite region between LINERs and star-forming galaxies. 

\section{Discussion}\label{sec:discussion}

\subsection{Occurrence and properties of \hii\ in the sample}

Of the 26 sources with usable \hi\ absorption spectra in our sample, we detected \hii\ in 5 sources. The detection rate is around 19\% (slightly higher if we exclude more targets  affected by RFI). Taking into account the uncertainties due to low number statistics, this fraction is, to first order, consistent with the results of \cite{Maccagni17}. 

We find that all the detections have unresolved radio continuum at the redshifted 21cm frequency at an angular resolution of \apx 3$''$ corresponding to \apx 20 kpc (Fig. \ref{fig:continuum_maps}), and all of them show extended features in VLBI images at parsec-scale resolution (see Fig. \ref{fig:radio_cont}). J1452+0627 and J0901+0304 are peaked-spectrum sources, which  suggests that they are young. J1458+3720 is classified as a blazar and has the least extended structure in the VLBI image. J2341+0018 is classified as a flat-spectrum source;   a classification is not available in the literature for J1429+0251. Furthermore, from the SDSS image, we find that J1452+0627 is a merger. 

The line widths (FWZI) of our detections range from \apx 60 \kmps\ in J1458+3720 to \apx 700 \kmps\ in J1429+0251. The \hii\ column densities we derive assuming a spin temperature of 100 K and a covering factor of unity range from $8 \times 10^{20}$ cm\p{2} to \apx 10\pp{22} cm\p{2}. As detailed in Sect. \ref{sec:hi_origin}, in J0901+0604 and J1452+0627 the absorption profile is consistent with \hii\ being distributed in a disc. Additionally, in the latter source we find that the \hii\ very likely has a high velocity-dispersion. In J1458+3720 the \hii\ clouds on galactic scales, farther away from the nuclear region, are likely to produce the observed absorption profile. In the case of J2341+0018 we find that a quiescent \hii\ disc can explain only a part of the observed profile. In the study of \citet{Maccagni17} it was found that broad and asymmetric \hi\ absorption profiles indicating disturbed \hii\ kinematics arose in higher power compact sources. However, we do not find extreme outflows of \hii\ despite the high radio power of the sources (see Sect. \ref{sec:modelling} for details).

\subsection{Comparison with other samples}

In the literature, there are about 60 other sources searched for \hi\ absorption in the redshift range $0.25 < z < 0.4$ with 11 detections of \hi\ absorption \citep{Carilli92, Pihlstrom03, Vermeulen03, Yan16, Aditya18a, Aditya18b, Gupta06, Curran19, Curran17, Curran06, Ostorero17, Curran11b, Grasha19}. Combining all these searches in the literature with our sample, we find an overall detection rate of 16\pc$^{+6\%}_{-5\%}$, again consistent within the errors with the detection rate at low redshifts. 

Likewise, these studies have also found that compact radio sources have a higher \hi\ detection rate in this redshift range, similar to what is seen at lower redshifts. Only the recent single-dish study of \citet{Grasha19} does not seem to agree with these results\footnote{Intriguingly, a few sources known to have \hi\ absorption, for example NGC\,1275 and 3C\,459 \citep[][respectively]{Morganti18b,Morganti01}, were not   detected by \citet{Grasha19}, although their data appear to be sensitive enough to detect them.}. In addition, a blind search for associated \hi\ absorption with the Australian SKA Pathfinder Array (ASKAP) at $0.34 < z < 0.79$ by \citet{Allison20} suggests that there is a difference between the \hi\ absorption detection rate at low redshifts \eg{Maccagni17} and the redshift range they cover. However, they agree that their non-detection of associated \hi\ absorption is consistent with the low-$z$ results if all the 14 sources in their sample are extended radio sources: their 3$\sigma$ optical depth limit may be much higher than reported in that case as the covering factor for extended radio sources is less than unity. Since at the spatial resolution of ASKAP it is not possible to ascertain the radio morphology of the sources in their sample, it is also not possible to test the difference in the detection rates between their sample and the low-$z$ studies. 

Earlier studies at much higher redshifts have proposed that the rest-frame UV luminosity of the AGN could affect the presence of \hii\ in radio AGN and hence the \hi\ detection rate, although whether and by how much the UV luminosity could affect the presence of \hii\ is still a matter of discussion. To explore this further we  investigated  the relation between the type of optical AGN and the detection of \hi\ absorption, and the direct relation to UV luminosity. To this end, we constructed the BPT diagram for our sample of objects. An analysis of the sample of \citet{Maccagni17} shows that there is no significant difference between the detection rate in the AGN falling under the category of LINERs and the powerful optical AGN falling in  the `Seyfert' region of the BPT diagram \citep{Santoro_thesis}. Our detections are also distributed among these two regions and we too do not see any skewed distribution of detections. If the optical AGN ionising the cold gas were   the reason for the \hi\ non-detections, then we would expect LINERs (where the optical AGN are not expected to be strong) to show a higher detection rate compared to the powerful optical AGN. We do not see such a trend in our study. However, our conclusions are limited by small number statistics. In any case, in our sample we find that most of the sources are below the proposed UV luminosity cutoff of 10\pp{23} W Hz\p{1} corresponding to the luminosity of unobscured AGN (see Fig. \ref{fig:uvlum}; \citealt{Curran08}).

A comparison of our \hi\ detection limit with the results of \hi\ emission studies could be insightful to understand the cause of non-detections. At low redshifts, \citet{Maccagni17} found that their $N_{\hii}$ detection limit in absorption, even after stacking all the non-detections, was higher than the typical \hii\ column density of the gas found in early-type galaxies. Though there is a study of stacking \hii\ in 21cm emission at the redshift range of our interest \citep{Bera19}, we find that their sample is dominated by blue star-forming galaxies, and they do not detect \hii\ by stacking red galaxies alone. Hence a  comparison similar to that done by \citet{Maccagni17} is not possible at the moment for our sample.

Overall, our findings hint towards the continuation of the trends proposed at lower redshifts by various studies \eg{Maccagni17, Gupta06, Vermeulen03}.

\subsection{The origin of \hi\ absorption}
\label{sec:hi_origin}

Detailed high spatial resolution studies have found the \hii\ seen in absorption in radio galaxies to exhibit a variety of structures: regularly rotating gas discs, high-velocity clouds, gas flowing outwards via direct interaction with radio jets, and gas falling towards the supermassive black hole (SMBH) \citep[e.g.][]{Struve10a,Morganti13, Schulz18, Schulz21}. Usually, symmetric \hi\ absorption profiles centred at or close to the systemic velocity, which are a few hundred \kmps wide, are attributed to rotating discs, while broader and asymmetric absorption profiles, blueshifted from the systemic velocity, are associated with gas outflows. Relatively narrow redshifted profiles sometimes arise from galactic clouds along the line of sight and sometimes from gas clouds falling into the SMBH. The kinematics of the absorbing gas can thus shed light on the nature of the interplay between the radio source and the surrounding ISM.

However, via \hi\ absorption we are only able to trace the gas present in front of the radio continuum. So the shape of the absorption profile also strongly depends on the morphology and the size of the radio continuum. 
For the detections reported in this paper the radio continuum appears to be \apx100 pc in size, and shows jet structures in all cases except J1458+3720  (see Fig. \ref{fig:radio_cont});  the absorption profiles exhibit multiple features in all the cases except J1458+3720.

Thus, the shape of the \hi\ absorption profile could be the result of a complex continuum structure even if the \hii\ is distributed in a regularly rotating disc. In order to ascertain the contribution to the absorption of a regularly rotating disc and gas with disturbed kinematics, we   performed a simple modelling for the three sources with the most complex \hi\ absorption profiles (J0901+0304, J1452+0627, and J2341+0018), as described in the next section. For the remaining two sources the spectra are either affected by RFI (J1429+0251), limiting the possibility of comparison with a model, or are very narrow (J1458+3720). For these sources a more qualitative approach was adopted.

The \hi\ absorption profile of J1429+0251 appears to be significantly broad with shallow wings and is symmetric and centred at the systemic velocity. The high-resolution VLBI image (Fig. \ref{fig:radio_cont}) shows the presence of two radio lobes. Since the absorption profile is mostly symmetric, it is very likely that the absorption arises from a gas disc seen against these lobes. However, since some of the line channels are affected by RFI the shape of the absorption profile is uncertain, and we cannot rule out the presence of any gas deviating from regular rotation.

The absorption profile of J1458+3720 is quite narrow (FWZI \apx 60 \kmps) and blueshifted by \apx 45 \kmps. The radio source has been classified as a blazar \citep{Massaro09}. Thus, the radio jets are expected to be close to the line of sight and perpendicular to the gas disc. Hence it is more likely that the absorption arises from gas clouds away from the nucleus, which usually show up as narrow absorption lines offset from the systemic velocity.

\begin{table}[]
\caption{Model parameters.}
\begin{tabularx}{\linewidth}{Xcccc}
\hline
Source & \Mstar & $v\rm_{rot}$ & Radio axis & Inclination\\
& ($10^{11}$ \Msun) & (\kmps) & $^\circ$ & $^\circ$ \\
(1) & (2) & (3) & (4) & (5)\\
\hline
J0901+0304  & 1.2  & 220  & 0  & $+$55\\
J1452+0627  & 2.6 & 300  & 80 & $-$40\\
J2341+0018  & 6.0  & 360    & 18 & $-$10\\
\hline
\end{tabularx}
\begin{tablenotes}
\item The columns are: (1) Radio source, (2) Stellar mass estimated as described in \ref{sec:modelling}, (3) Rotation velocity (4) Position angle of the radio source, (5) Inclination angle used to obtain the models shown in Fig. \ref{fig:rotcur}.
\end{tablenotes}
\label{model_param}
\end{table}

\begin{figure}[h!]
    \includegraphics[width=\linewidth]{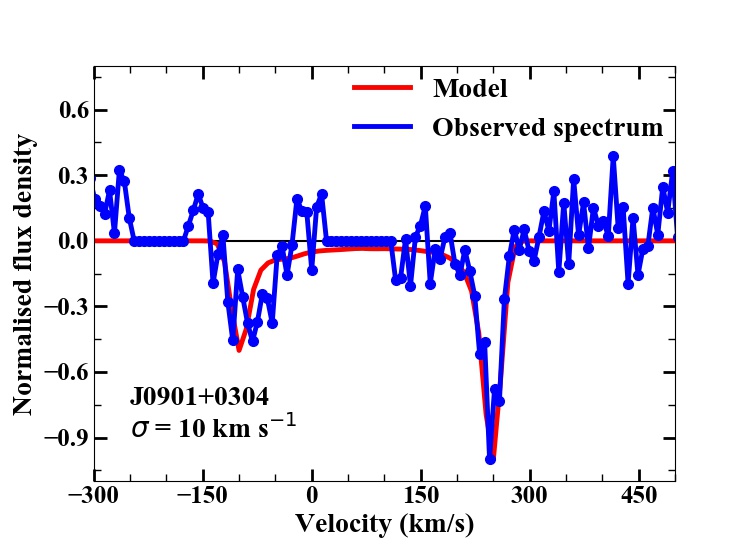}
    \includegraphics[width=\linewidth]{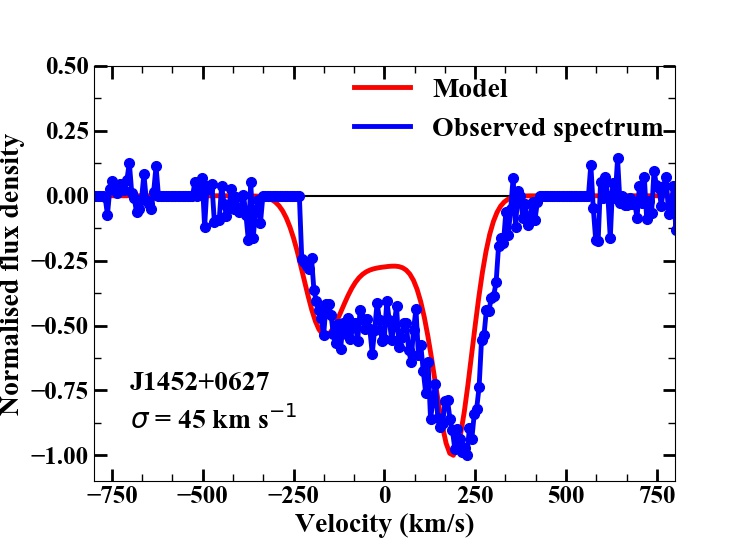}
    \includegraphics[width=\linewidth]{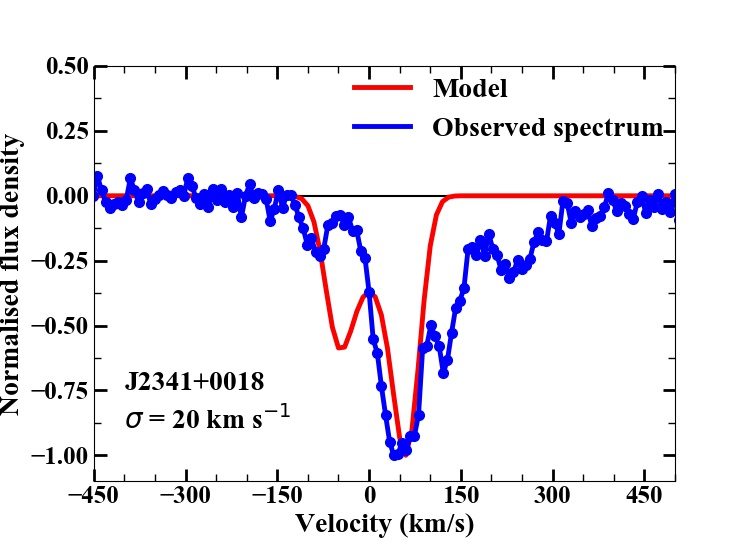}
    \caption{Model \hi\ absorption profiles for J0901+0304, J1452+0627, and J2341+0018 overlaid on the observed spectra. The velocity dispersion of the gas is mentioned in the figures. The model profile of J0901+0304 is shifted by 70 \kmps\ to show that the relative strengths of the two components and their separation are well reproduced by the model. See Sect. \ref{sec:modelling} for more details.}
    \label{fig:rotcur}
\end{figure}

\subsection{Possible \hii\ discs in J0901+0304, J1452+0627, and J2341+0018}
\label{sec:modelling}

We modelled the \hi\ absorption in these three cases to assess whether for the given radio continuum the profiles can indeed arise from a regularly rotating disc. 

The shape of the absorption profile depends strongly on the shape of this background continuum structure. Since the radio sources are unresolved in our observations, we need higher resolution radio images that clearly resolve them. At the redshifted \hii\ 21cm frequency, there are no observations available that clearly resolve the radio structure. However, as mentioned in the earlier sections, VLBI images at the desirable spatial resolution are available at much higher radio frequencies. Thus, for our purposes we  used the VLBI X-band images (8 GHz), which are the highest spatial resolution images available (see Fig. \ref{fig:radio_cont}). This entails an assumption that the radio continuum structure as seen at 8 GHz is similar to that at the redshifted \hii\ 21cm frequency of \apx 1 GHz. It is also quite possible that an additional diffuse kiloparsec-scale radio emission is present at the redshifted \hii\ 21cm frequency that is missed by higher frequency radio observations. However, that is unlikely to be the major continuum component giving rise to absorption. As discussed in Section 4.2, the high-frequency high-resolution image of J2341+0018 does indeed represent the dominant part of the radio continuum. We do not have enough spectral information available for J0901+0304 and J1452+0627 to draw a similar conclusion. However, as we show below, the X-band continuum structure can already  almost entirely reproduce the observed absorption profile in both the cases. 

We  used the centre of the radio sources  shown in Fig. \ref{fig:radio_cont} as  the location of the core, and hence the centre of the \hii\ disc being modelled, except in the case of J0901+0304. Here we   assumed a core to be slightly to the south of the centre shown in Fig. \ref{fig:radio_cont} (more details below). We note that a spectral-index study will be needed to confirm the location of the core in all these cases.

The parameters involved in this modelling are the velocity dispersion of gas, the inclination and the position angles, and the rotation curve of the galaxy. The sources under consideration have SDSS photometry available. We estimated the stellar masses using the correlation between the K-corrected g$-$r colour and the mass-to-light ratio \citep{Bell03}. We then used the baryonic Tully--Fisher relation \citep{McGaugh00} to estimate the rotation velocity. Furthermore, in the absence of detailed studies available for the host galaxies, we assumed that the rotation curve is flat in each case. The stellar masses obtained and the corresponding rotation velocities are tabulated in Table \ref{model_param}. 

The other important parameters are the position angle (PA) and the inclination angle of the disc. We considered \hii\ discs that are parallel to the radio axis. We note that if the disc is perpendicular to the radio axis, the main radio structure would be aligned with the minor axis of the disc which would result in a narrow profile centred on the systemic velocity, unlike the observed complex profiles. In the case of J1452+0627 the HST image  shows that the dust lane is aligned   with the radio source, supporting our assumption. This suggested alignment of \hii\ discs and radio axes in our detections is intriguing since radio jets are usually found to be perpendicular to the gas discs \citep[][and references therein]{Morganti18}. To obtain the model absorption profiles, we only varied the inclination angle by hand. The PA of the radio source used, and hence that of the model \hii\ disc and the inclination angle,  are also given in Table. \ref{model_param}.

The absorption profile of J0901+0304 consists of two components, one redshifted and the other blueshifted with respect to the systemic velocity. \citet{Yan16} report a slightly different redshift (see Sect. \ref{sec:0901_results}) according to which the shallow component is at the systemic velocity and the narrow deep component is redshifted. Based on this finding, they hypothesise that the shallow absorption feature arises from an \hii\ disc, while the redshifted narrow profile could be due to an infalling high-velocity cloud. However, we find that  using the SDSS DR13 redshift value shows that the absorption profile essentially straddles the systemic velocity, indicating that the \hii\ detected in absorption is more likely in a disc. Given the apparent uncertainty in the redshift of this source, we  applied our model to verify how well a disc structure could reproduce the absorption.

The model absorption profile for J0901+0304, shown in Fig. \ref{fig:rotcur}, was obtained for a radio core that is slightly to the south of the centre shown in Fig. \ref{fig:radio_cont}. We chose this to be the location of the radio core since the observed double-horn profile was reproduced only for a radio source consisting of a core and two radio jets. The profile we obtain is quite similar to the observed spectrum, with the same velocity separation between the two components and also their relative strengths. However, the profile is offset towards bluer velocities by \apx70 \kmps\ compared to the observed profile. We note that this could be due to uncertainties on the redshift of J0901+0304. However, the remarkable similarity in the shape of the profile suggests that the absorption is indeed very likely from a disc.

For J1452+0627, which is a merger, the absorption profile we obtain matches the observed spectrum quite well. However, we find that the velocity dispersion needed to explain the profile is quite high, \apx 45 \kmps. In comparison, typical  \hii\ discs are found to have a velocity dispersion of \apx 10 \kmps\ \citep[e.g.][]{Struve10c}. A lower velocity dispersion in our model gives rise to narrower profiles that do not match the observed spectrum. Our model thus suggests that the gas seen in absorption is distributed in a turbulent gas disc quite similar to that seen in B2\,0258+35 \citep{Murthy19}. The optical emission lines for this source consist of two kinematic components, both wide and one blueshifted by \apx 300 \kmps, indicating a significant disturbance in the nuclear region of this galaxy. This further supports the suggested high velocity-dispersion of \hii.

For J2341+0018, our model reproduces only the deep absorption component at the systemic velocity. The blueshifted component we obtain is deeper than the observed blueshifted wing. We were not able to obtain the redshifted feature for any inclination angle. Although redshifted absorption profiles indicating infalling gas clouds are not uncommon, they  tend to be narrower and deeper compared to what is observed here. We  note that our model consists of a single regularly rotating \hii\ disc. Any warp in the disc, which our model does not account for, might very well give rise to such features. Thus, we conclude that  an \hii\ disc alone is not able to explain the asymmetries in the observed profile in the case of J2341+0018.

We conclude that the observed absorption profiles in J0901+0304 and J1452+0624 are consistent with the \hii\ arising from regularly rotating discs (a disc with high velocity dispersion in the latter case). In the case of J2341+0018, a quiescent \hii\ disc can only explain the deep absorption component, while it does not reproduce the shallow redshifted and the blueshifted wings. Intriguingly, at least for the three objects where the modelling has been done, the profiles are explained only if the \hii\ discs are aligned with (or close to) the position angle of the radio (VLBI) continuum emission. This is surprising since circumnuclear discs are often suggested and found to be perpendicular to the radio axis (e.g. Cygnus A by \citealt{Struve10a} and Centaurus A \citealt{Struve10b}).

\section{Summary}
\label{sec:summary}

We have carried out a pilot search for \hi\ absorption in a sample of 30 radio-loud AGN with the JVLA-A array expanding the study at low redshifts by \citet{Gereb15} and \citet{Maccagni17}. Our sample spans the redshift range from 0.25 to 0.4. Of the 26 sources making up our final sample, in our observations, at spatial scales of \apx 5-15 kpc, the radio continua of 12 sources are unresolved, while the other 14 are resolved. We detect \hi\ absorption in five sources, which  gives an \hi\ detection rate of \apx 19\%. This agrees with the detection rate obtained by combining our sample with the sources studied in the same redshift range available in the literature. All five of our detections are in compact radio sources and one of these is also a merger. Thus, our results expand the result at low redshift providing a comparable detection rate, albeit with a large error bar due to the small number statistics. 

The absorption profiles we detect range from being quite narrow (FWZI \apx60 \kmps) to very wide (FWZI \apx700 \kmps), and are  mostly of a complex nature. The \hii\ column densities range from 8 $\times 10^{20}$ cm\p{2} to 10\pp{22} cm\p{2}. The radio sources in our sample are more powerful than those typically studied at low redshifts. We also find that the optical ionisation lines in about half the sample exhibit broad blueshifted kinematic components, indicating a significant disturbance in the nuclear region. However, despite the high radio power, many of them are low-ionisation sources. We also find that the UV luminosities of most of our sources and all our detections are below the proposed threshold above which all the cold gas is supposed to have been entirely ionised. 

An analysis of the \hii\ absorption spectra shows that in two cases the observed profile is consistent with the \hii\ distributed in a gas disc. In one more case we find that a quiescent disc cannot explain the asymmetries in the profile, while in another \hii\ clouds on galactic scales away from the nuclear region could be responsible for the observed profile. The ionised gas shows multiple kinematic components in some of the detections further supporting the presence of disturbed gas. We cannot ascertain the morphology of the absorbing gas in one other case as quite a few line channels are affected by RFI. However, given the symmetry and the width of the line in this case \citep[FWZI \apx 700 \kmps; similar to that seen in high-power sources at low redshifts;][]{Maccagni17}, it is likely that the absorption, in this case, arises from a combination of a gas disc and disturbed gas. 

Since our findings suggest that it is very likely that the detection rate of \hii\ is similar (and hence quite high), even at the redshift range covered, the next step is to expand the sample size to include more sources of different radio powers and covering different regions in various parameter spaces like the WISE colour-colour and the BPT diagrams. This will allow us to carry out a more detailed analysis of the interplay between \hii\ and the radio source in different classes of AGN at these redshifts. The ongoing and upcoming \hi\ surveys will also contribute in this direction, although they will need follow-up at subarcsecond resolutions to ascertain the continuum structure of the AGN and the morphology of the absorbing gas. 

\begin{acknowledgements}

We are grateful to the referee for a thorough read of the manuscript and the useful comments. Part of the research leading to these results has received funding from the European Research Council under the European Union's Seventh Framework Programme (FP/2007-2013) / ERC Advanced Grant RADIOLIFE-320745. The National Radio Astronomy Observatory is a facility of the National Science Foundation operated under cooperative agreement by Associated Universities, Inc. The National Radio Astronomy Observatory is a facility of the National Science Foundation operated under cooperative agreement by Associated Universities, Inc. We used the Astrogeo VLBI FITS image database publicly available at http://astrogeo.org/vlbi\_images. We have made use of the ``K-corrections calculator'' service available at http://kcor.sai.msu.ru/. Funding for the Sloan Digital Sky Survey (SDSS) has been provided by the Alfred P. Sloan Foundation, the Participating Institutions, the National Aeronautics and Space Administration, the National Science Foundation, the U.S. Department of Energy, the Japanese Monbukagakusho, and the Max Planck Society. The SDSS Web site is http://www.sdss.org/. The SDSS is managed by the Astrophysical Research Consortium (ARC) for the Participating Institutions. The Participating Institutions are The University of Chicago, Fermilab, the Institute for Advanced Study, the Japan Participation Group, The Johns Hopkins University, Los Alamos National Laboratory, the Max-Planck-Institute for Astronomy (MPIA), the Max-Planck-Institute for Astrophysics (MPA), New Mexico State University, University of Pittsburgh, Princeton University, the United States Naval Observatory, and the University of Washington. 

\end{acknowledgements}

%
   \bibliographystyle{aa} 
   \bibliography{ref} 
%


\begin{appendix}

\section{Additional tables}

\begin{landscape}
\begin{table}[]
\caption{Observations details}
\begin{tabular}{ccccccccccccc}
\hline
Source & \textit{z} & $\nu \rm_{obs}$ & $\rm \Delta t$ & BW & $\rm \Delta$v & Beam$_{\rm robust}$ & PA$_{\rm robust}$ & RMS$\rm_{map,robust}$ & Beam$_{\rm natural}$ & PA$_{\rm natural}$ & RMS$_{\rm map,natural}$ & RMS$\rm_{cube}$\\ 
 & & (MHz) & (mins) & (MHz) & (\kmps) & ($^{\prime\prime} \times ^{\prime\prime})$ & ($^{\circ}$) & ($\mu$Jy beam\p{1}) & ($^{\prime\prime} \times ^{\prime\prime}$) & ($^{\circ}$) & ($\mu$Jy beam\p{1}) & (\mjypb) \\ 
 (1) & (2) & (3) & (4) & (5) & (6) & (7) & (8) & (9) & (10) & (11) & (12) & (13)\\
\hline

SDSS\,J003058+011600 &   0.3748      & 1033.17 & 30 & 32 & 7.25 & 2.53 $\times$  1.93 & -67.71 & 300.00 & 2.53 $\times$  1.93 & -49.57 & 270 & 4.7 \\
SDSS\,J075622+355442 &   0.3191      & 1076.75 & 30 & 32 & 6.96 & 0.95 $\times$  0.95 &  45.00 & 301.49 & 2.34 $\times$  2.28 &  42.90 & 273 & 3.0 \\
SDSS\,J083825+371037 &   0.3962      & 1017.26 & 40 & 32 & 7.37 & 2.10 $\times$  1.60 & -73.73 & 567.79 & 2.75 $\times$  2.34 &  81.14 & 2000& 2.6 \\
SDSS\,J090151+030423 &   0.2875      & 1103.18 & 40 & 32 & 6.79 & 2.82 $\times$  1.72 &  60.28 & 220.94 & 4.11 $\times$  2.15 &  52.45 & 570 & 2.6 \\
SDSS\,J102844+384437 &   0.3621      & 1042.71 & 40 & 32 & 7.19 & 2.11 $\times$  1.59 & -73.98 & 270.42 & 2.94 $\times$  2.45 &  75.81 & 510 & 2.4 \\
SDSS\,J110655+190912 &   0.3400      & 1059.87 & 30 & 32 & 7.05 & 2.08 $\times$  1.62 &  79.61 & 282.88 & 2.94 $\times$  2.49 & -58.55 & 330 & 3.4 \\
SDSS\,J112342+125214 &   0.3178      & 1077.88 & 30 & 32 & 6.95 & 0.95 $\times$  0.95 &  45.00 & 231.13 & 2.37 $\times$  2.04 & -39.23 & 350 & 3.2 \\
SDSS\,J112833+324323 &   0.3695      & 1037.11 & 30 & 32 & 7.23 & 1.87 $\times$  1.49 &  48.50 & 705.32 & 2.46 $\times$  2.29 &  44.08 & 1700& 3.9 \\
SDSS\,J114409+005436 &   0.2955      & 1096.33 & 30 & 32 & 6.84 & 1.99 $\times$  1.54 &  59.14 & 328.38 & 2.58 $\times$  2.07 &   4.59 & 500 & 4.0 \\
SDSS\,J115324+493109 &   0.3339      & 1092.83 & 40 & 32 & 6.85 & 1.54 $\times$  1.34 &  16.32 & 241.32 & 1.96 $\times$  1.76 &  32.65 & 260 & 6.0 \\
SDSS\,J123413+475351 &   0.3736      & 1034.06 & 40 & 32 & 7.25 & 1.84 $\times$  1.63 &  16.13 & 822.12 & 2.61 $\times$  2.40 &  47.22 & 1300& 2.5 \\
SDSS\,J140416+411749 &   0.3604      & 1044.09 & 30 & 32 & 7.17 & 3.18 $\times$  1.77 & -89.97 & 405.29 & 4.30 $\times$  2.71 &  78.25 & 1000& 2.5 \\
SDSS\,J140854+302103 &   0.2897      & 1101.29 & 30 & 32 & 6.80 & 3.83 $\times$  1.44 &  72.94 & 171.41 & 5.15 $\times$  2.04 &  65.62 & 200 & 2.6 \\
SDSS\,J142904+025140 &   0.2931      & 1098.44 & 40 & 32 & 8.82 & 1.87 $\times$  1.58 &  69.21 & 194.97 & 2.51 $\times$  2.08 & -16.77 & 460 & 2.3 \\
SDSS\,J145239+062738 &   0.2671      & 1120.93 & 40 & 32 & 6.68 & 1.67 $\times$  1.54 & -80.35 & 109.12 & 2.54 $\times$  2.19 &   0.70 & 400 & 3.1 \\
SDSS\,J145338+035933 &   0.3697      & 1036.81 & 30 & 32 & 7.22 & 2.61 $\times$  2.05 &  62.77 & 203.01 & 4.03 $\times$  2.62 &  49.73 & 450 & 2.3 \\
SDSS\,J145845+372022 &   0.3332      & 1065.28 & 30 & 32 & 7.03 & 1.92 $\times$  1.68 &  66.10 & 239.28 & 4.92 $\times$  2.90 &  52.45 & 385 & 3.0 \\
SDSS\,J145859+041614 &   0.3916      & 1020.66 & 40 & 32 & 7.34 & 3.93 $\times$  1.46 &  78.10 & 130.72 & 2.69 $\times$  2.24 &  69.38 & 670 & 3.2 \\
SDSS\,J150720+364245 &   0.3837      & 1026.53 & 30 & 32 & 7.29 & 1.36 $\times$  1.36 &  45.00 & 248.06 & 4.55 $\times$  2.67 &  73.55 & 700 & 2.5 \\
SDSS\,J152503+110744 &   0.3326      & 1065.86 & 30 & 32 & 7.03 & 3.13 $\times$  1.85 &  68.22 & 414.76 & 4.45 $\times$  2.42 &  58.27 & 1350& 2.5 \\
SDSS\,J153617+462734 &   0.3696      & 1037.09 & 30 & 32 & 7.22 & 3.55 $\times$  1.56 &  81.58 & 182.38 & 4.99 $\times$  2.65 &  76.01 & 600 & 2.0 \\
SDSS\,J162214+090421 &   0.2917      & 1099.61 & 30 & 32 & 6.82 & 3.43 $\times$  1.81 &  62.60 & 215.21 & 5.29 $\times$  2.38 &  56.72 & 340 & 2.5 \\
SDSS\,J164734+270558 &   0.2822      & 1107.86 & 30 & 32 & 6.70 & 1.91 $\times$  1.54 & -76.26 & 179.17 & 3.33 $\times$  2.45 &  70.37 & 370 & 2.0 \\
SDSS\,J165547+232931 &   0.3931      & 1019.53 & 30 & 32 & 7.35 & 1.87 $\times$  1.69 & -69.72 & 161.10 & 2.66 $\times$  2.52 &  57.55 & 180 & 2.7 \\
SDSS\,J172109+354216 &   0.2832      & 1106.87 & 30 & 32 & 6.70 & 1.93 $\times$  1.51 & -67.95 & 243.45 & 2.66 $\times$  2.32 &  87.85 & 950 & 2.0 \\
SDSS\,J234107+001833 &   0.2767      & 1112.54 & 30 & 32 & 6.70 & 2.37 $\times$  1.79 & -78.35 & 454.57 & 3.33 $\times$  2.45 & -47.97 & 2500& 2.2 \\
SDSS\,J085451+621850 &   0.2673      & 1120.74 & RFI&    &      &         \\ 
SDSS\,J111141+355337 &   0.3048      & 1088.60 & RFI&    &      &          \\
SDSS\,J114539+442022 &.  0.2997      & 1065.05 & RFI&    &      &         \\

\hline
\end{tabular}
\begin{tablenotes}
\item The columns are: (1) Source name, (2) SDSS DR13 redshift, (3) redshifted \hi\ line frequency in MHz, (4) On-source time in minutes, (5) Bandwidth of the observations in MHz, (6) Spectral resolution in \kmps, (7), (8) and (9) Beam size, position angle and RMS of the continuum image made with ROBUST $-1$ weighting (10), (11) and (12) Beam size, position angle and RMS of the continuum image made with natural weighting, (13) RMS noise on the cube made with natural weighting with the same beam parameters as the naturally weighted continuum image.
\end{tablenotes}
\label{radio_obs}
\end{table}
\end{landscape}

\small
\begin{landscape}
\begin{table}[]
\caption{The radio properties of the AGN detected in \hi\ absorption.}
\begin{tabular}{ccccccccccccccc}
\hline
Source & $z$ & S$_{\nu}$ (int) & S$_{\nu}$ (peak) & $\rm \Delta v$ & RMS$\rm_{spec}$ & $\tau\rm_{peak}$& $\int(\tau$ dv)  & N$\rm_{HI}$ & L$\rm_{1.4 GHz}$ & L$\rm_{UV}$ \\
 &  & (mJy) & (mJy beam\p{1}) & (\kmps) &  (mJy beam\p{1}) & (\%) & (\kmps)  & $\times 10^{20}$ (cm\p{2}) & (W Hz\p{1}) & (W Hz\p{1})\\ 
 (1) & (2) & (3) & (4) & (5) & (6) & (7) & (8) & (9) & (10) & (11)\\
\hline
SDSS\,J090151+030423 & 0.2875    & 380.0&   371.0  &  6.79  &  2.6  &    7.0   &  4.53 $\pm$   0.30   &    8.25  $\pm$   0.55 &  26.0    &      21.4 \\
SDSS\,J142904+025140 & 0.2931    & 845.0&   810.0  &  8.82  &  2.3  &    3.0   &  5.59 $\pm$   0.31   &    10.19  $\pm$  0.60 &  26.32   &      18.8 \\
SDSS\,J145239+062738 & 0.2671    & 300.0&   228.4  &  6.68  &  3.1  &    20.7  &  72.68 $\pm$   1.35   &   132.49 $\pm$  2.47 &  25.6    &      20.4 \\
SDSS\,J145845+372022 & 0.3332    & 194.0&   183.0  &  7.03  &  3.0  &    33.7  &  6.72  $\pm$   0.62   &   12.26  $\pm$ 1.13  &  25.6    &      21.2 \\
SDSS\,J234107+001833 & 0.2767    & 443.1&   429.5  &  6.7   &  2.2  &    6.4   &  9.64  $\pm$   0.33   &   17.58 $\pm$  0.65  &  25.9    &      20.6 \\
\hline
\end{tabular}
\begin{tablenotes}
\item The columns are: (1) Source name, (2) SDSS DR13 redshift, (3) Integrated radio continuum flux in mJy, (4) Peak flux density in \mjypb, (5) Velocity resolution of the \hi\ spectra in \kmps, (6) RMS noise on the spectrum at the velocity resolution listed in (5) in \mjypb, (7) Peak optical depth (8) Integrated optical depth in \kmps (9) \hii\ column density in units of 10$^{20}$ cm\p{2} estimated assuming a \tspin $\rm = 100$ K and $c_f =1$, (10) Rest-frame 1.4GHz radio luminosity in log scale, (11) Rest-frame UV luminosity in log scale.
\end{tablenotes}
\label{radio_table_det}
\end{table}
\end{landscape}

\begin{landscape}
\begin{table}[]
\caption{Derived parameters for the \hi\ non-detections.}
\begin{tabular}{cccccccccccc}
\hline
Source   & $z$  & S$\rm_{\nu}$ (int) & S$\rm_{\nu}$ (peak)  & $\rm \Delta s$    & $\tau$  & N$\rm_{HI}$  & L$\rm_{1.4 GHz}$ & L$\rm_{UV}$  & Continuum      \\
& & (mJy) & (mJy beam\p{1}) &  (mJy beam\p{1}) &  & (cm\p{2}) & W Hz\p{1} & W Hz\p{1} & morphology\\
(1) & (2) & (3) & (4) & (5) & (6) & (7) & (8) & (9) & (10)\\
\hline
SDSS\,J003058+011600  &   0.3748     &   416.3  &   189.0  &     1.0    &   0.017  &   20.22  &   25.92  &   -      &   e\\
SDSS\,J075622+355442  &   0.3191     &   506.25 &   130.0  &     0.7    &   0.017  &   20.21  &   25.6   &   22.83  &   e\\
SDSS\,J083825+371037  &   0.3962     &   482.2  &   442.4  &     0.6    &   0.004  &   19.63  &   26.34  &   21.00  &   c\\
SDSS\,J102844+384437  &   0.3621     &   787.0  &   600.0  &     0.7    &   0.003  &   19.53  &   26.39  &   20.74  &   e\\
SDSS\,J110655+190912  &   0.3400     &   970.0  &   190.0  &     1.0    &   0.016  &   20.2   &   25.83  &   19.81  &   e\\
SDSS\,J112342+125214  &   0.3178     &   273.2  &   204.6  &     1.1    &   0.016  &   20.19  &   25.8   &   20.98  &   e\\
SDSS\,J112833+324323  &   0.3695     &   393.0  &   371.3  &     1.5    &   0.012  &   20.08  &   26.2   &   20.03  &   c\\
SDSS\,J114409+005436  &   0.2955     &   335.0  &   350.0  &     1.0    &   0.008  &   19.92  &   25.97  &   21.47  &   c\\
SDSS\,J115324+493109  &   0.3339     &   2060.0 &   1000.0 &     0.9    &   0.002  &   19.43  &   26.53  &   23.20  &   e\\
SDSS\,J123413+475351  &   0.3736     &   372.4  &   353.2  &     0.6    &   0.005  &   19.73  &   26.19  &   22.73  &   c\\
SDSS\,J140416+411749  &   0.3604     &   300.0  &   272.5  &     1.0    &   0.012  &   20.06  &   26.04  &   20.9   &   c\\
SDSS\,J140854+302103  &   0.2897     &   393.0  &   235.4  &     0.6    &   0.008  &   19.89  &   25.77  &   20.44  &   e\\
SDSS\,J145338+035933  &   0.3697     &   475.0  &   421.5  &     0.7    &   0.005  &   19.69  &   26.25  &   22.34  &   e\\
SDSS\,J145859+041614  &   0.3916     &   940.0  &   250.0  &     0.9    &   0.011  &   20.02  &   26.08  &   21.67  &   e\\
SDSS\,J150720+364245  &   0.3837     &   539.0  &   482.0  &     0.9    &   0.005  &   19.75  &   26.35  &   22.89  &   e\\
SDSS\,J152503+110744  &   0.3326     &   537.6  &   512.2  &     0.6    &   0.004  &   19.59  &   26.24  &   -      &   c\\
SDSS\,J153617+462734  &   0.3696     &   308.7  &   198.4  &     0.4    &   0.007  &   19.86  &   25.93  &   20.86  &   e\\
SDSS\,J162214+090421  &   0.2917     &   356.3  &   256.5  &     0.5    &   0.006  &   19.81  &   25.81  &   20.49  &   e\\
SDSS\,J164734+270558  &   0.2822     &   650.0  &   470.0  &     0.5    &   0.003  &   19.51  &   26.04  &   21.10  &   e\\
SDSS\,J165547+232931  &   0.3931     &   318.7  &   312.3  &     0.7    &   0.006  &   19.82  &   26.18  &   20.6   &   c\\
SDSS\,J172109+354216  &   0.2832     &   228.4  &   212.11 &     0.6    &   0.008  &   19.91  &   25.7   &   22.59  &   c\\
SDSS\,J085451+621850  &   0.2673     &   RFI \\
SDSS\,J111141+355337  &   0.3048     &   RFI \\
SDSS\,J114539+442022  &   0.2997     &   RFI \\
\hline
\end{tabular}
\begin{tablenotes}
\item The columns are: (1) Source name, (2) SDSS DR13 redshift, (3) Integrated radio continuum flux in mJy, (4) Peak flux density in \mjypb, (5) RMS noise on the spectrum in \mjypb after Hanning smoothing it to a spectral resolution close to 50 \kmps, (6) 3$\sigma$ optical depth limit, (7) 3$\sigma$ limit on the \hii\ column density in log scale estimated assuming a \tspin $\rm= 100$ K and c$_f =1$, (8) Rest-frame 1.4GHz radio luminosity in log scale, (9) Rest-frame UV luminosity in log scale, (10) Continuum morphology as seen in our observations: e is extended; c is compact. Sources with the ratio S$_\nu$(peak)/S$_\nu$(int)$> 0.9$ were considered compact and the rest extended sources. The UV luminosity was estimated using the SDSS u- and g-band magnitudes. The values are left blank where the photometry was not available.
\end{tablenotes}
\label{radio_table_nondet}
\end{table}
\end{landscape}

\begin{table*}[]
\caption{The optical line fluxes and the line ratios derived from the SDSS spectra.}
\begin{tabular}{cccccccccc}
\hline
Source    & $z$     & \hii & Kinematic & [OIII]   &  [NII] & H$\alpha$  & H$\beta$ & [OIII]/H$\beta$ & [NII]/H$\alpha$  \\
& & flag & components & flux & flux & flux & flux\\
(1) & (2) & (3) & (4) & (5) & (6) & (7) & (8) & (9) & (10)\\
\hline
SDSS\,J003058+011600   &   0.3748       &   0   &   0   &      -       &   -          &   -         &   -               &  -                 &  -        \\
SDSS\,J075622+355442   &   0.3191       &   0   &   0   &      -       &   -          &   -         &   -               &  -                 &  -        \\ 
SDSS\,J083825+371037   &   0.3962       &   0   &   1   &      33.07   &   -          &   -         &   7.55            &  0.64              &  -        \\
SDSS\,J085451+621850   &   0.2673       &   0   &   1   &      -       &   76.83      &   55.15     &   -               &  -                 &  0.14     \\         
SDSS\,J090151+030423   &   0.2875       &   1   &   2   &      27.26   &   100.31     &   145.71    &   21.45           &  0.10              & -0.16     \\         
SDSS\,J102844+384437   &   0.3621       &   0   &   2   &      95.00   &   151.11     &   95.23     &   26.25           &  0.55              &  0.20     \\        
SDSS\,J110655+190912   &   0.3400       &   0   &   1   &      30.07   &   -          &   -         &   5.18            &  0.76              &  -        \\
SDSS\,J111141+355337   &   0.3048       &   0   &   2   &      235.52  &   18.53      &   69.21     &   42.26           &  0.74              &  -0.5     \\          
SDSS\,J112342+125214   &   0.3178       &   0   &   2   &      186.66  &   -          &   -         &   18.48           &  1.00              &  -        \\
SDSS\,J112833+324323   &   0.3695       &   0   &   0   &      -       &   -          &   -         &   -               &  -                 &  -        \\  
SDSS\,J114409+005436   &   0.2955       &   0   &   1   &      52.57   &   105.29     &   41.22     &   4.05            &  1.11              &  0.40     \\        
SDSS\,J114539+442022   &   0.2997       &   0   &   2   &      96.36   &   111.45     &   29.96     &   15.17           &  0.80              &  0.57     \\         
SDSS\,J115324+493109   &   0.3339       &   0   &   2   &      577.55  &   1357.01    &   1013.28   &   1018.82         & -0.24              &   0.12     \\         
SDSS\,J123413+475351   &   0.3736       &   0   &   1   &      249.85  &   3508.60    &   273.73    &   1111.90         & -0.64              &   1.10     \\        
SDSS\,J140416+411749   &   0.3604       &   0   &   1   &      54.87   &   -          &   -         &   12.92           &  0.62              &  -        \\ 
SDSS\,J140854+302103   &   0.2897       &   0   &   2   &      57.15   &   -          &   -         &   11.48           &  0.69              &  -        \\
SDSS\,J142904+025140   &   0.2931       &   0   &   1   &      24.90   &   51.58      &   46.70     &   8.31            &  0.47              &  0.04     \\         
SDSS\,J145239+062738   &   0.2671       &   1   &   2   &      366.76  &   382.93     &   339.60    &   25.07           &  1.16              &  0.05     \\         
SDSS\,J145338+035933   &   0.3697       &   0   &   1   &      -       &   72.97      &   28.24     &   -               &  -                 &  0.41     \\         
SDSS\,J145859+041614   &   0.3916       &   0   &   2   &      159.13  &   112.11     &   132.13    &   23.73           &  0.82              & -0.07     \\         
SDSS\,J145845+372022   &   0.3332       &   1   &   0   &      -       &   -          &   -         &   4.42            &  -                 &  -        \\
SDSS\,J150720+364245   &   0.3837       &   0   &   1   &      30.068  &   81.83      &   31.48     &   4.35            &  0.83              &  0.41     \\        
SDSS\,J152503+110744   &   0.3326       &   0   &   2   &      341.84  &   231.15     &   220.52    &   41.74           &  0.91              &  0.02     \\         
SDSS\,J153617+462734   &   0.3696       &   0   &   2   &      403.69  &   95.64      &   163.60    &   52.27           &  0.88              & -0.23     \\         
SDSS\,J162214+090421   &   0.2917       &   0   &   0   &      -       &   -          &   -         &   -               &  -                 &  -        \\
SDSS\,J164734+270558   &   0.2822       &   0   &   0   &      -       &   -          &   -         &   -               &  -                 &  -        \\
SDSS\,J165547+232931   &   0.3931       &   0   &   2   &      209.02  &   74.18      &   107.25    &   36.69           &  0.75              & -0.16     \\         
SDSS\,J172109+354216   &   0.2832       &   0   &   2   &      961.08  &   -          &   -         &   89.27           &  1.03              &  -        \\
SDSS\,J234107+001833   &   0.2767       &   1   &   1   &      146.74  &   692.62     &   810.68    &   140.26          &  0.01              & -0.06     \\        
\hline
\end{tabular}
\begin{tablenotes}
\item The columns are: (1) Source name, (2) SDSS DR13 redshift, (3) \hi\ detection flag: 0 $=$ non-detection, 1$=$detection, (4) Number of kinematic components (from the \oiii\ line), (5) \oiii\ flux, (6) [N II]$\lambda\lambda$6548,84$\rm \AA$ flux, (7) \halpha\ flux, (8) H$\beta$ flux, (9) log$_{10}$([OIII]/H$\beta$), (10) log$_{10}$([NII]/\halpha). In case of multiple kinematic components, the flux listed is the total flux of all the components.
\end{tablenotes}
\label{optical_table}
\end{table*}

\section{HI spectra of the non-detections}
\label{sec:HIundetected}
\begin{figure*}[!h]
\includegraphics[width=6cm, height=4cm]{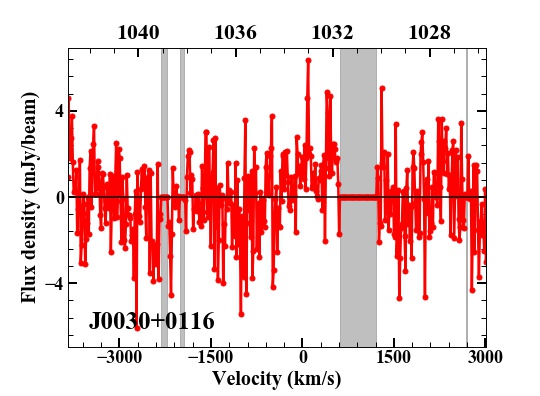}
\includegraphics[width=6cm, height=4cm]{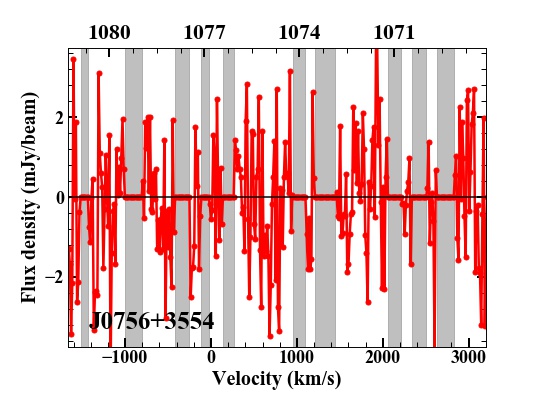}
\includegraphics[width=6cm, height=4cm]{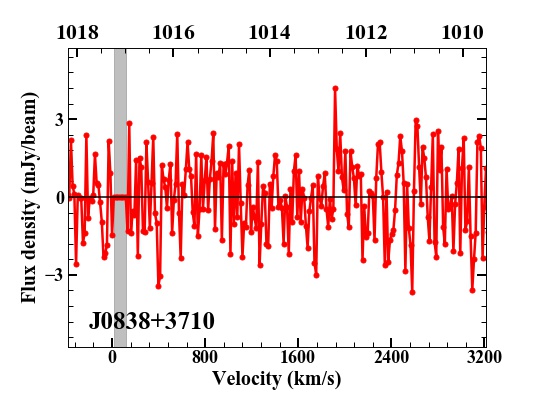}
\includegraphics[width=6cm, height=4cm]{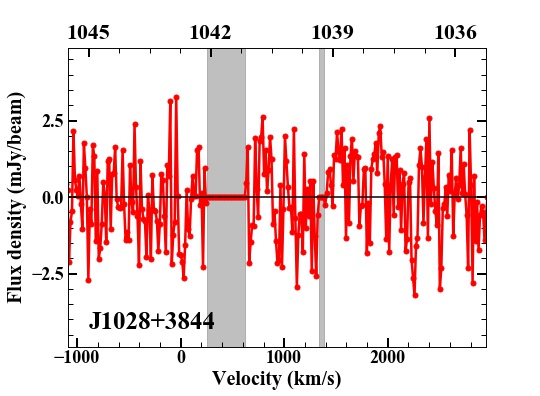}
\includegraphics[width=6cm, height=4cm]{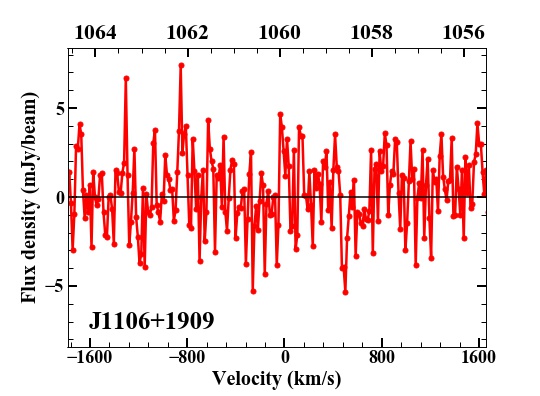}
\includegraphics[width=6cm, height=4cm]{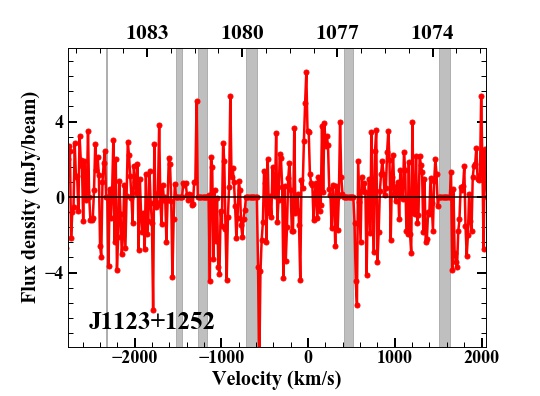}
\includegraphics[width=6cm, height=4cm]{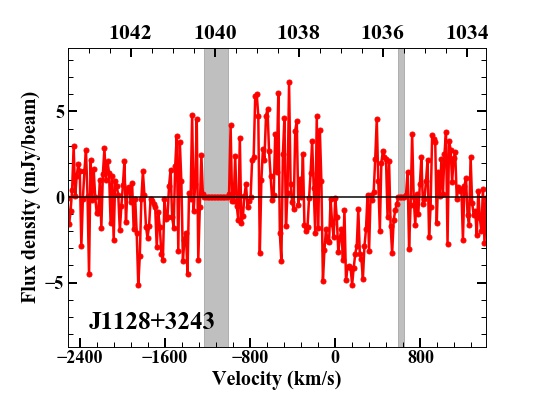}
\includegraphics[width=6cm, height=4cm]{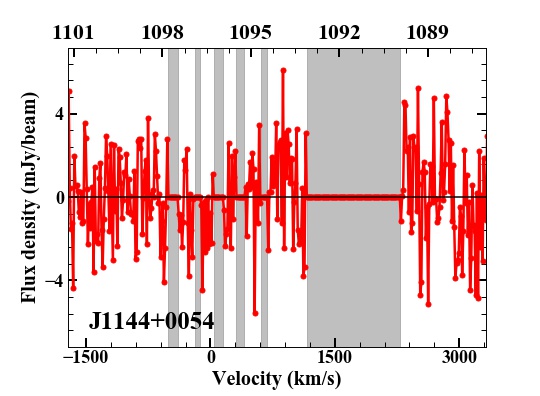}
\includegraphics[width=6cm, height=4cm]{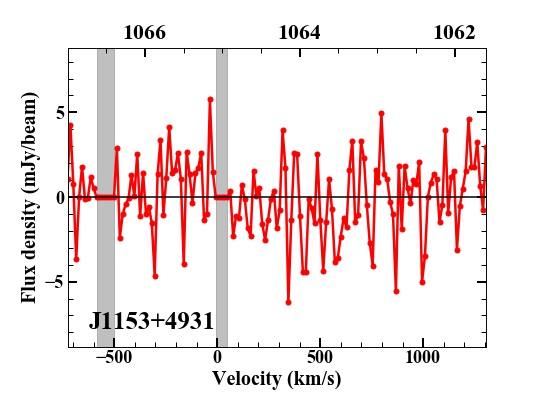}
\includegraphics[width=6cm, height=4cm]{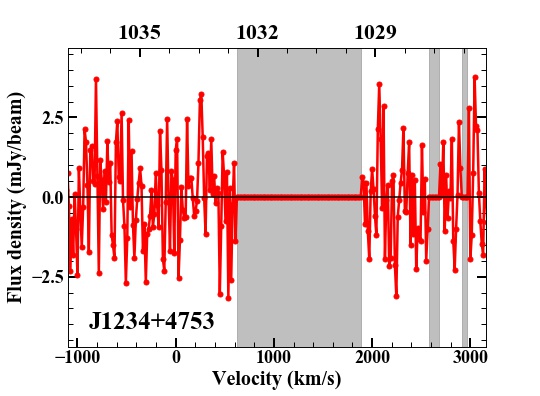}
\includegraphics[width=6cm, height=4cm]{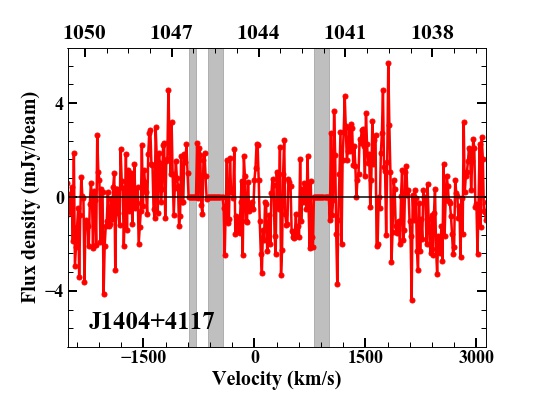}
\includegraphics[width=6cm, height=4cm]{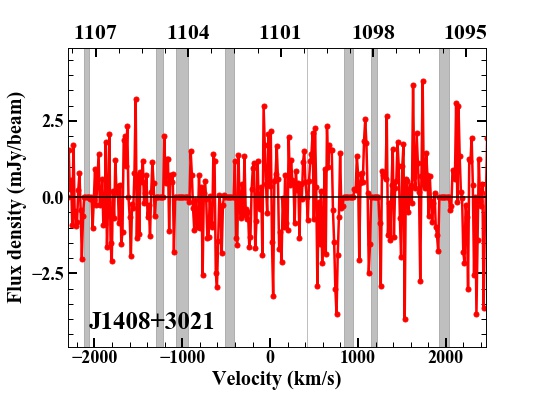}
\includegraphics[width=6cm, height=4cm]{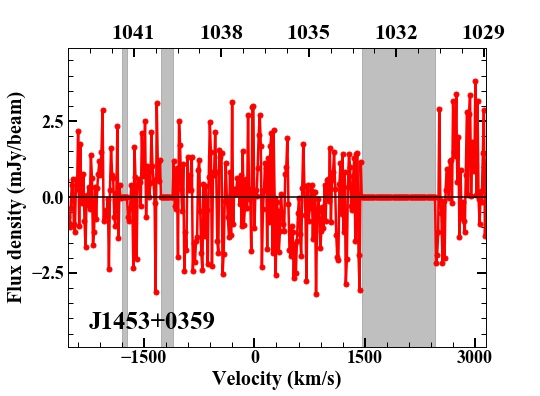}
\includegraphics[width=6cm, height=4cm]{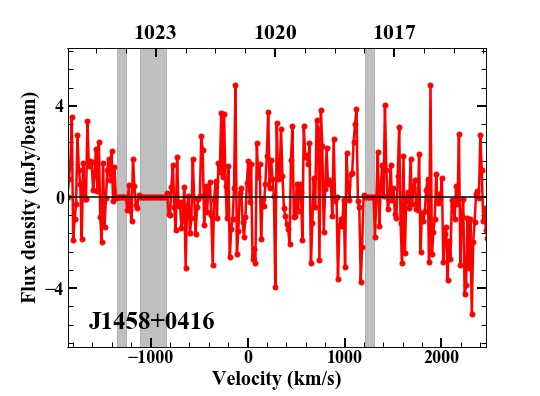}
\includegraphics[width=6cm, height=4cm]{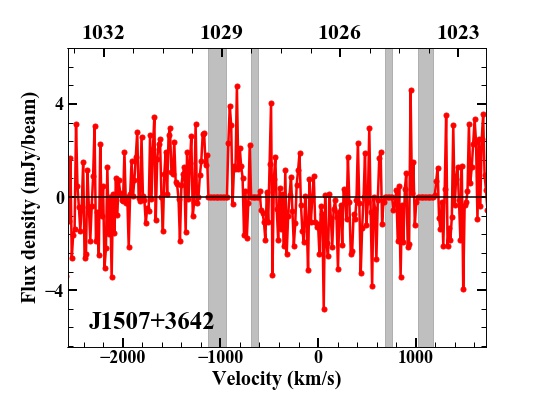}
\includegraphics[width=6cm, height=4cm]{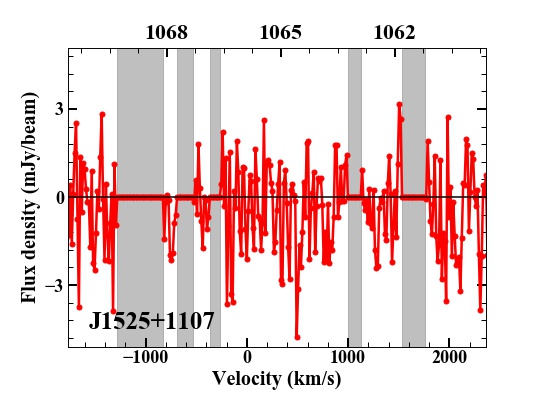}
\hspace{0.1em}
\includegraphics[width=6cm, height=4cm]{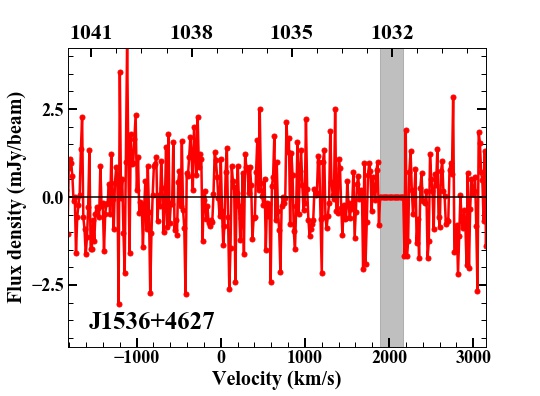}
\hspace{0.1em}
\includegraphics[width=6cm, height=4cm]{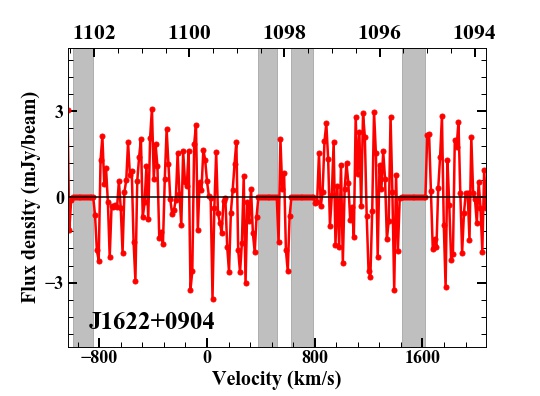}
   \caption{Spectra of the non-detections. The heliocentric frequency, in MHz, is indicated at the top of each panel. The grey shaded region shows the velocity ranges affected by RFI.}
    \label{fig:nondet_spectra}
\end{figure*}

\begin{figure*}\ContinuedFloat
\includegraphics[width=6cm, height=4cm]{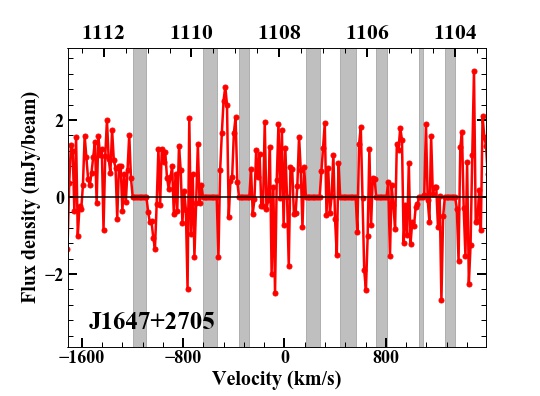}
\includegraphics[width=6cm, height=4cm]{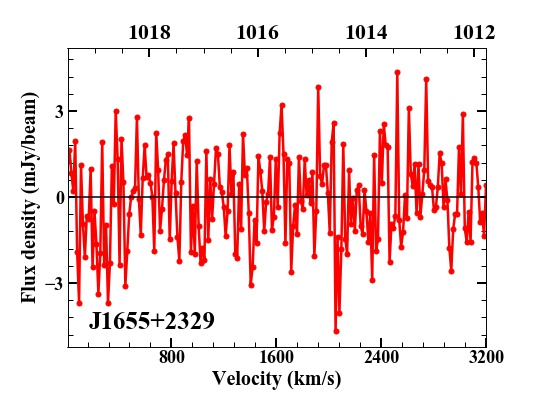}
\hspace{0.6em}
\includegraphics[width=6cm, height=4cm]{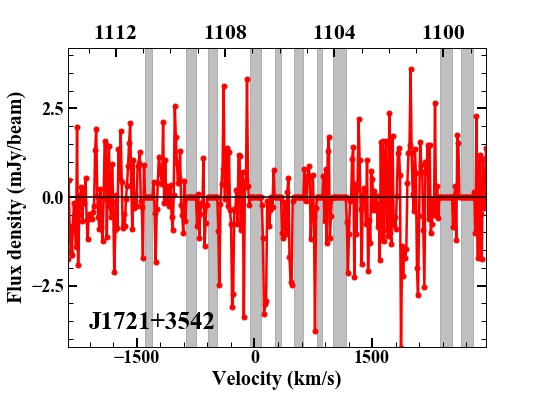}
   \caption{Continued.}
    \label{fig:nondet_spectra}
\end{figure*}

\section{Continuum images of the radio sources}
\label{sec:ContImages}

\begin{figure*}

\includegraphics[width=6cm, height=4.3cm]{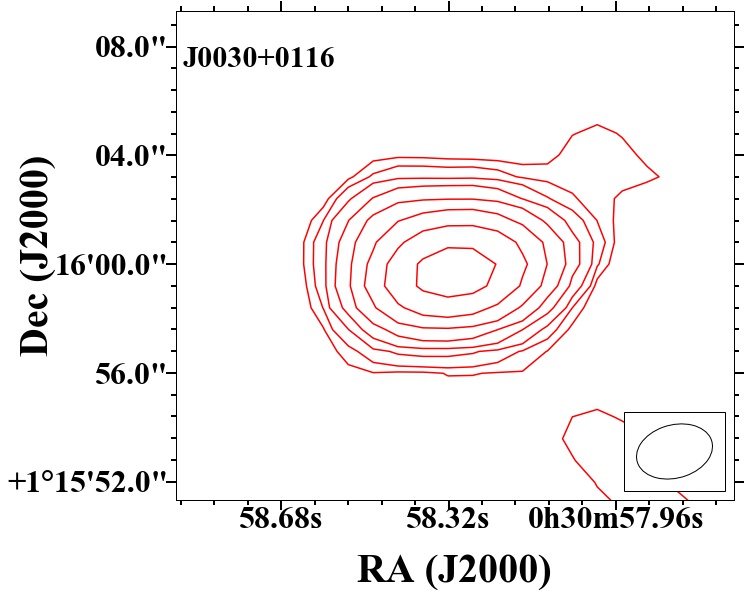}
\includegraphics[width=6cm, height=4.3cm]{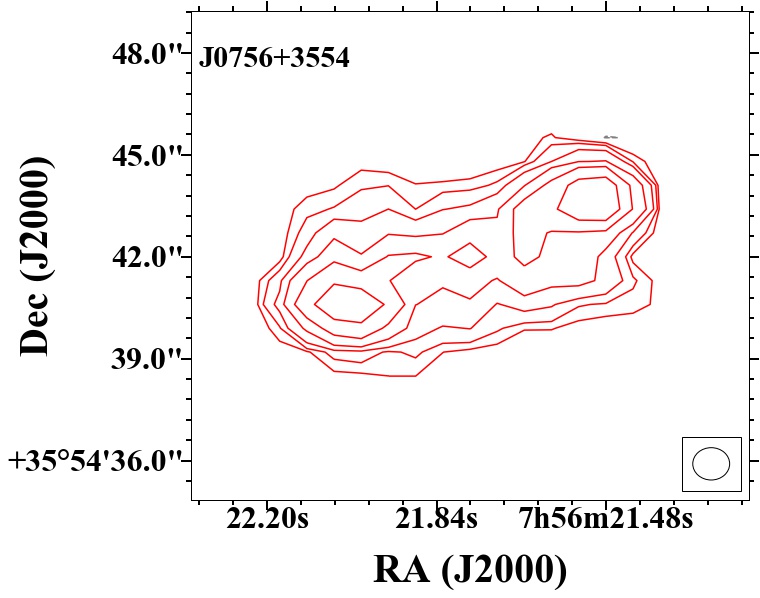}
\includegraphics[width=6cm, height=4.3cm]{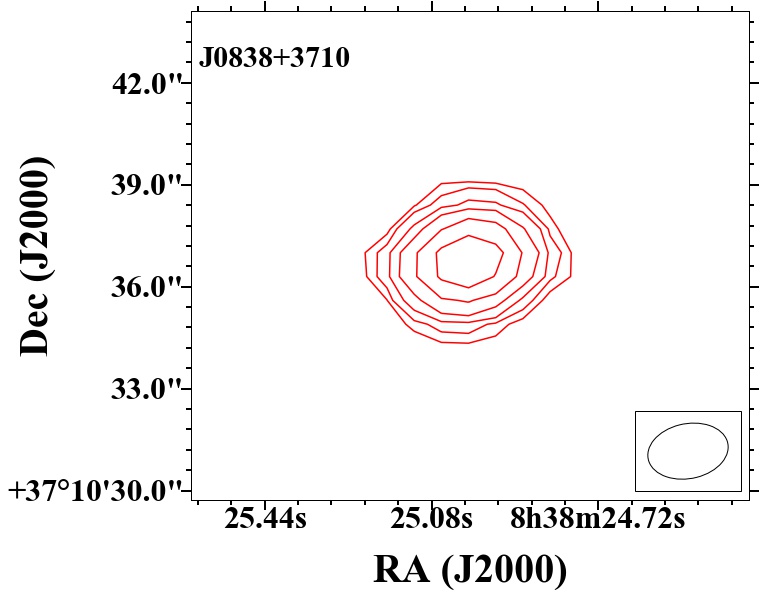}
\includegraphics[width=6cm, height=4.3cm]{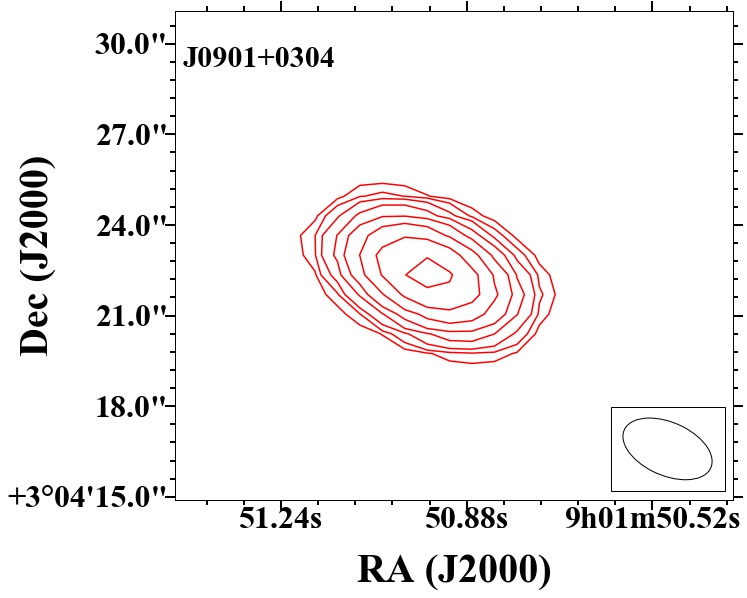}
\includegraphics[width=6cm, height=4.3cm]{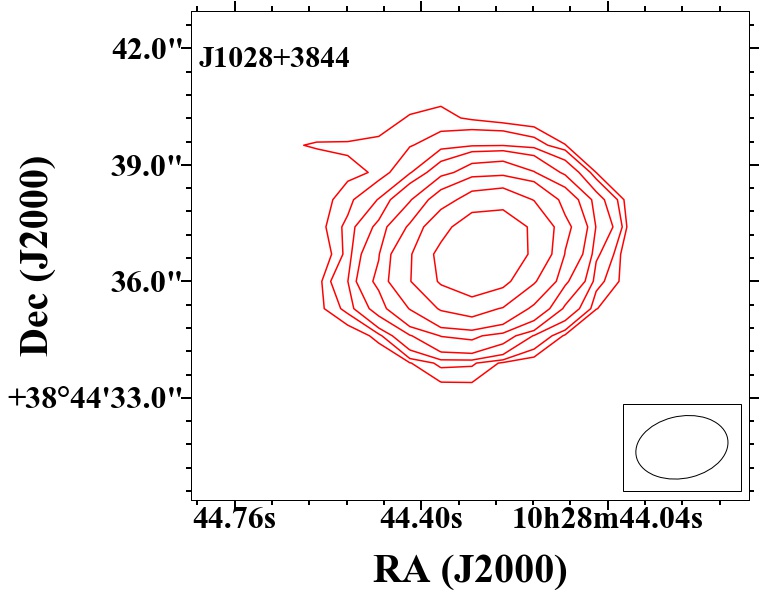}
\includegraphics[width=6cm, height=4.3cm]{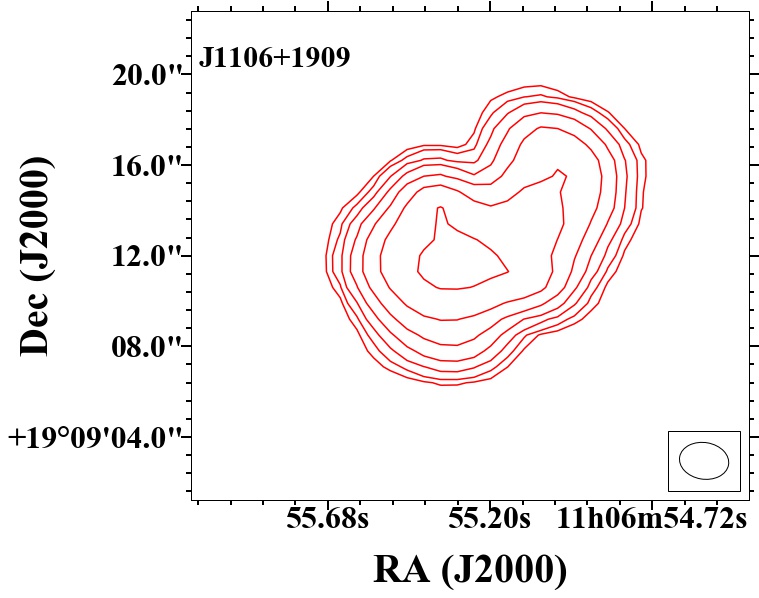}
\includegraphics[width=6cm, height=4.3cm]{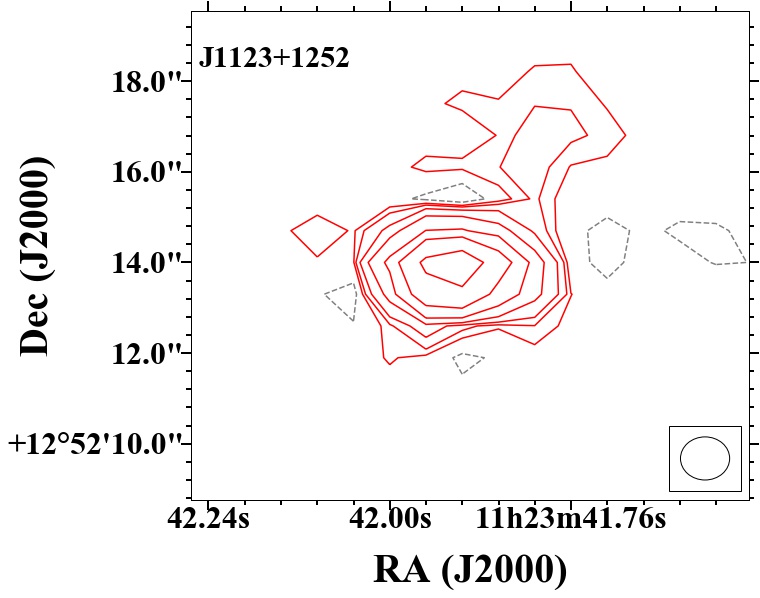}
\includegraphics[width=6cm, height=4.3cm]{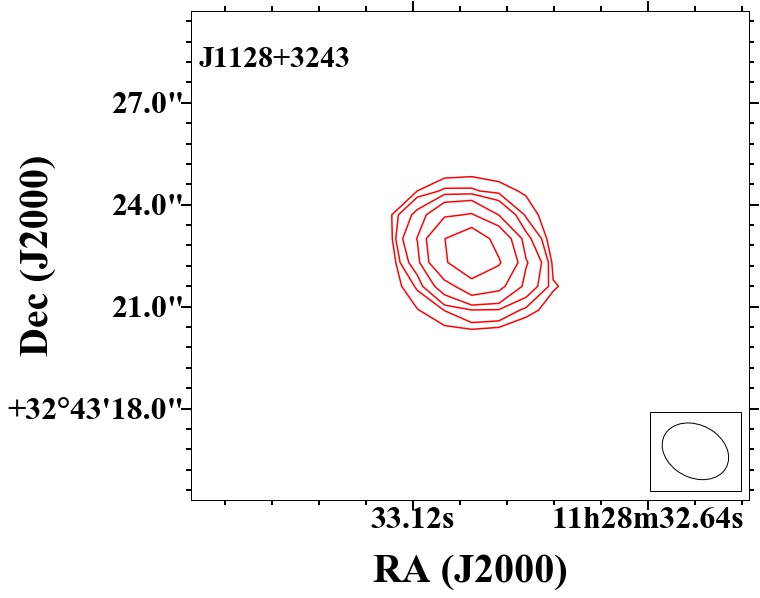}
\includegraphics[width=6cm, height=4.3cm]{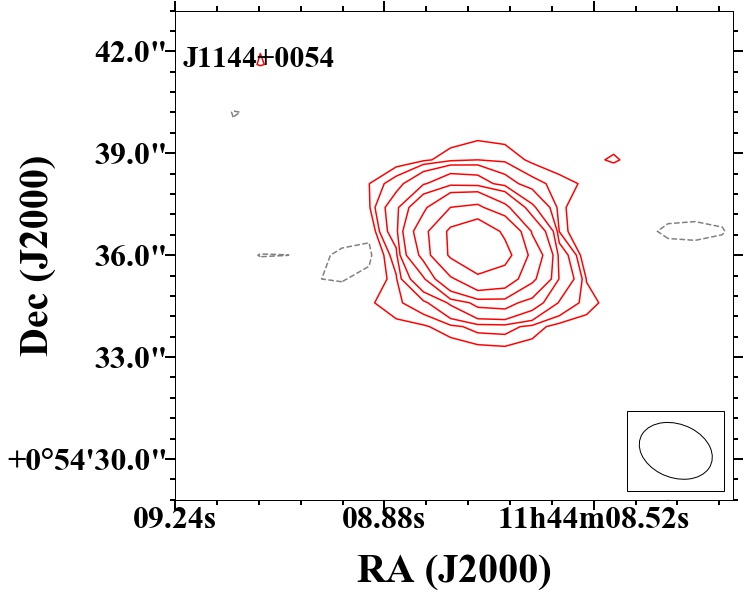}
\includegraphics[width=6cm, height=4.3cm]{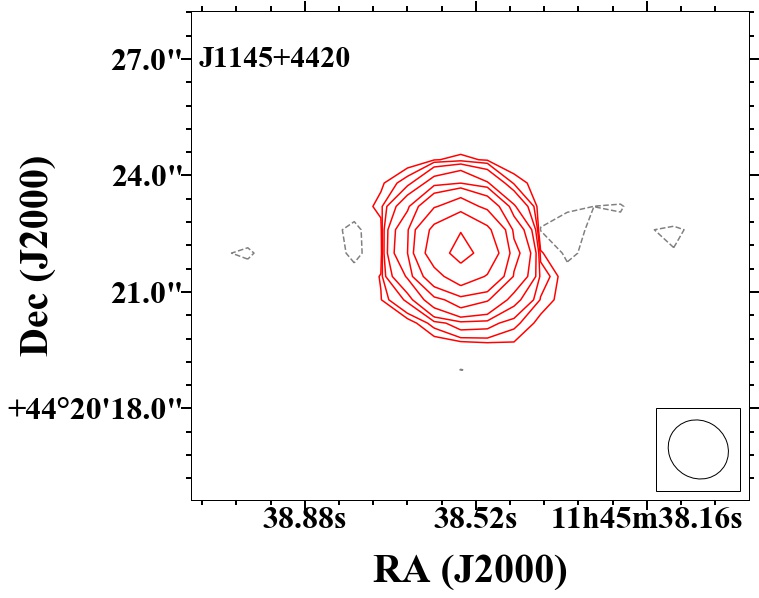}
\includegraphics[width=6cm, height=4.3cm]{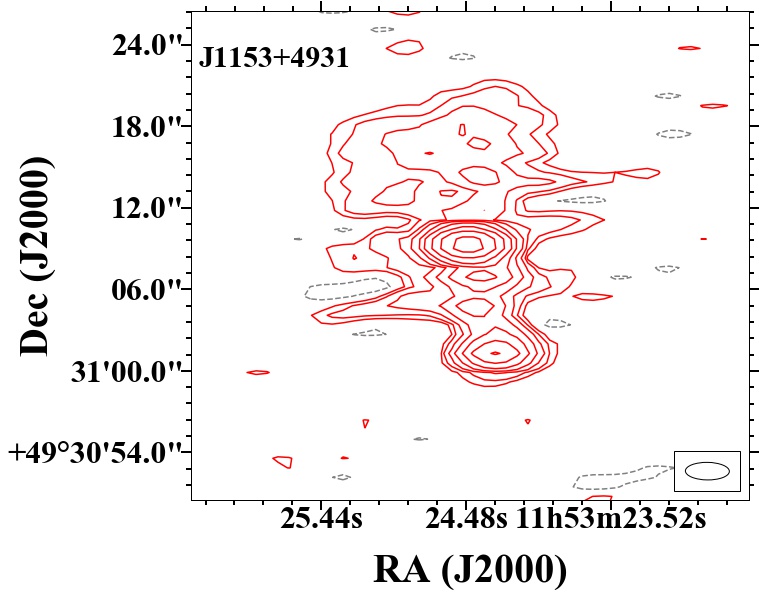}
\includegraphics[width=6cm, height=4.3cm]{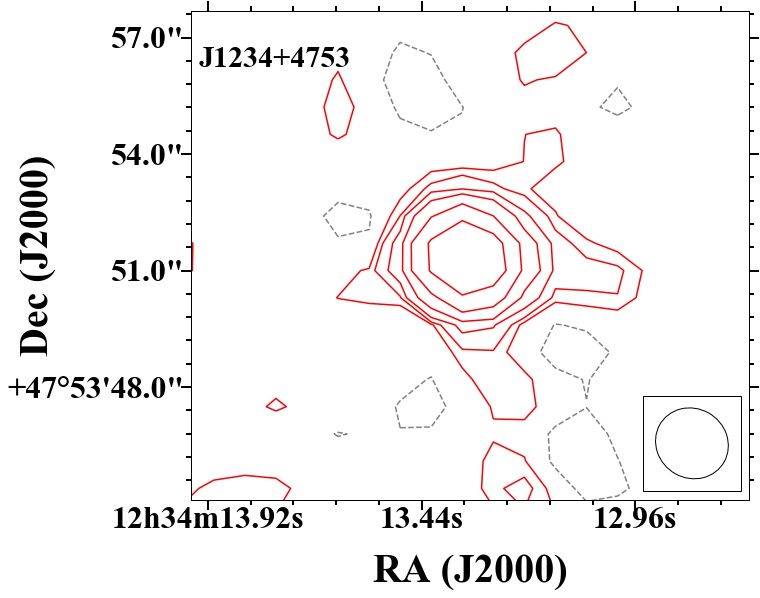}
\includegraphics[width=6cm, height=4.3cm]{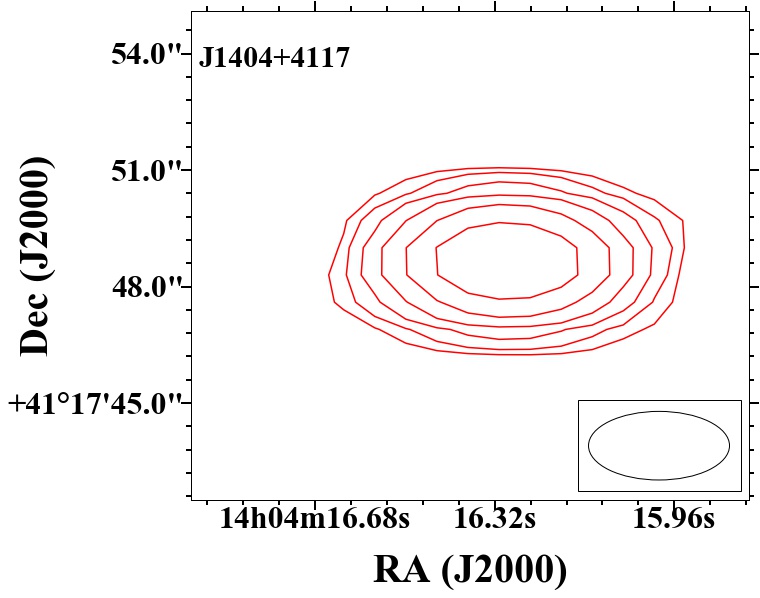}
\hspace{0.1cm}
\includegraphics[width=6cm, height=4.3cm]{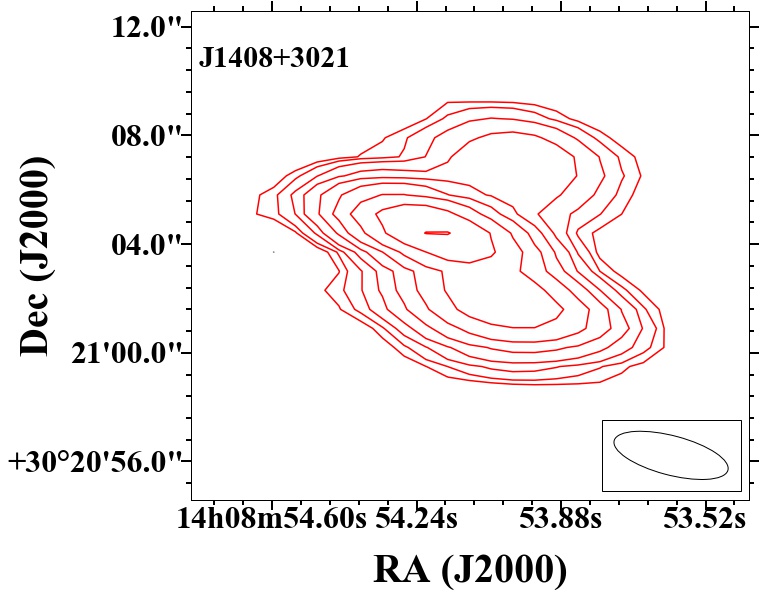}
\hspace{0.1cm}
\includegraphics[width=6cm, height=4.3cm]{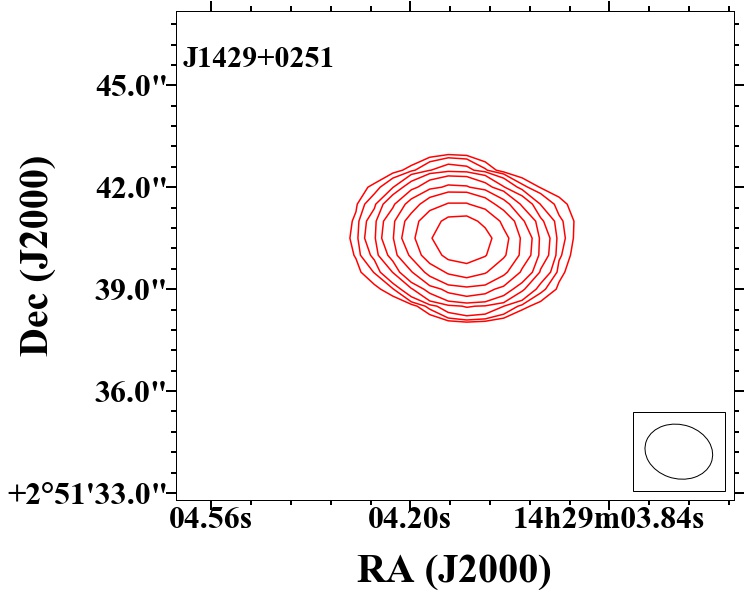}

\caption{Radio continuum images of the sources in our sample. The contours start at 4$\sigma$ and increase by steps of two. The RMS noise and the beam sizes are listed in Table \ref{radio_table_nondet}. The 4$\sigma$ negative contours are shown in grey.}
\label{fig:continuum_maps}
\end{figure*}

\begin{figure*}
\ContinuedFloat
\includegraphics[width=6cm, height=4.3cm]{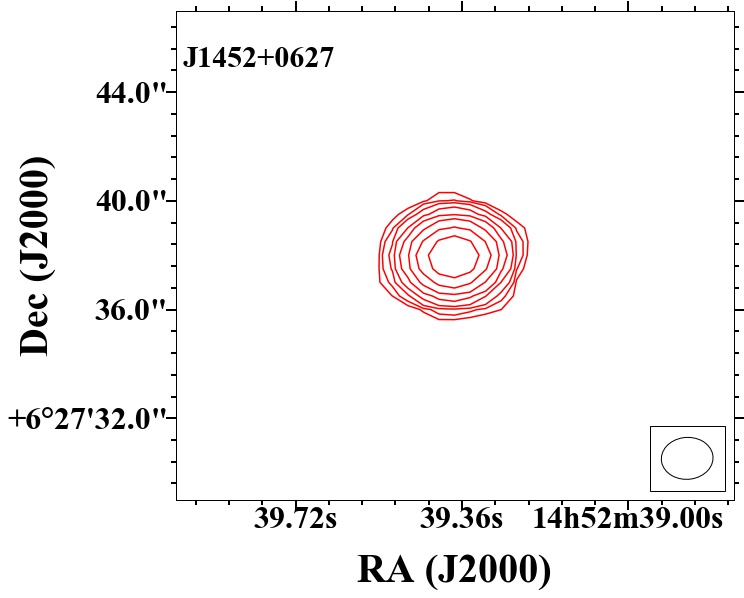}
\includegraphics[width=6cm, height=4.3cm]{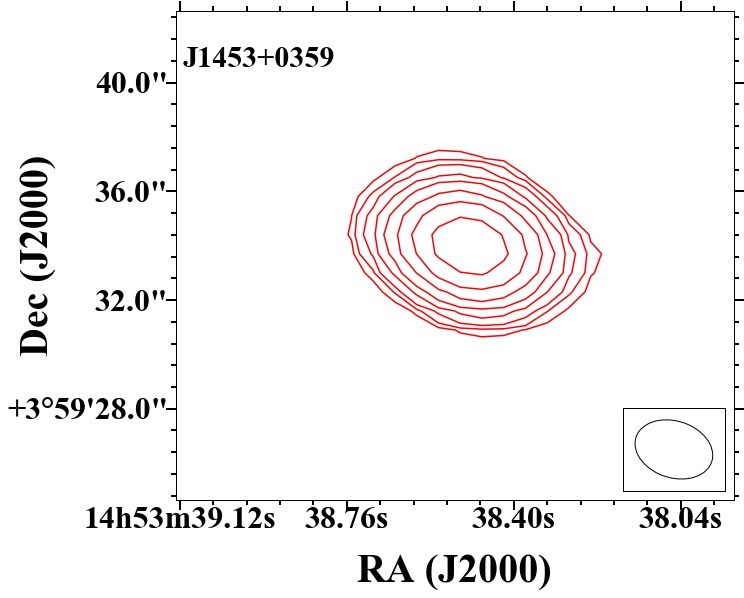}
\includegraphics[width=6cm, height=4.3cm]{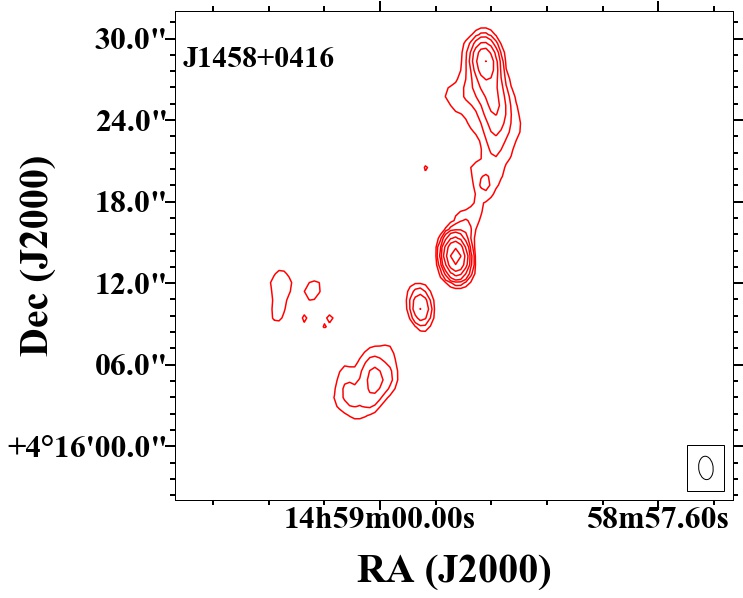}
\includegraphics[width=6cm, height=4.3cm]{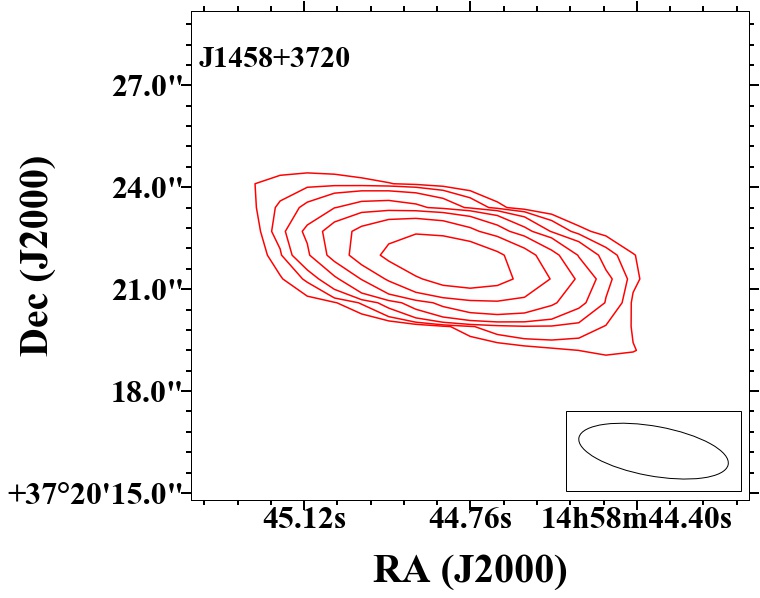}
\includegraphics[width=6cm, height=4.3cm]{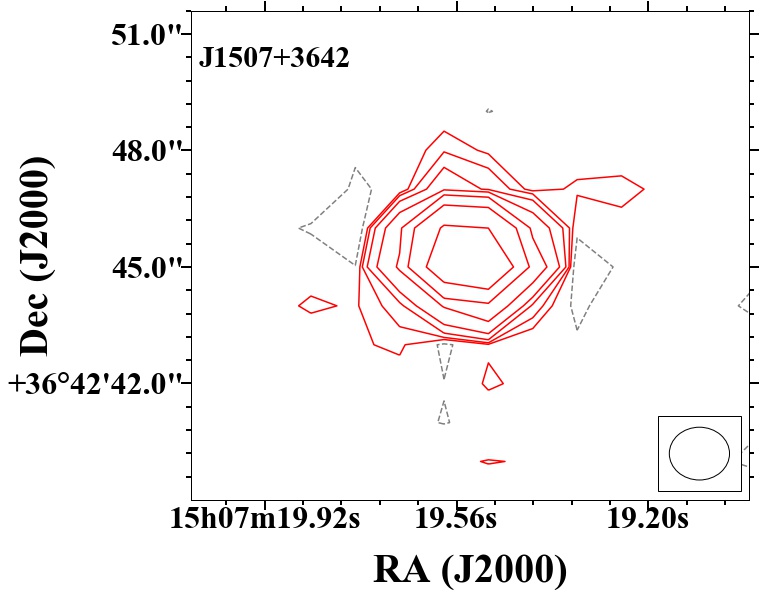}
\includegraphics[width=6cm, height=4.3cm]{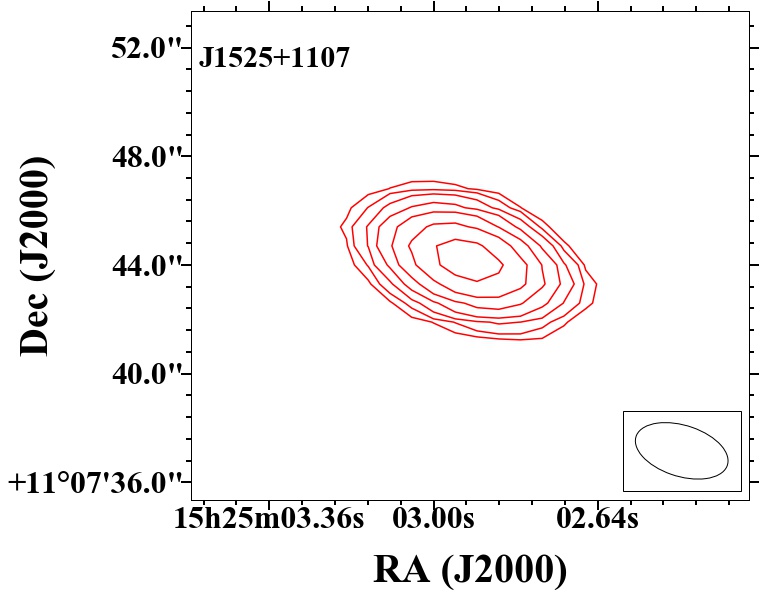}
\includegraphics[width=6cm, height=4.3cm]{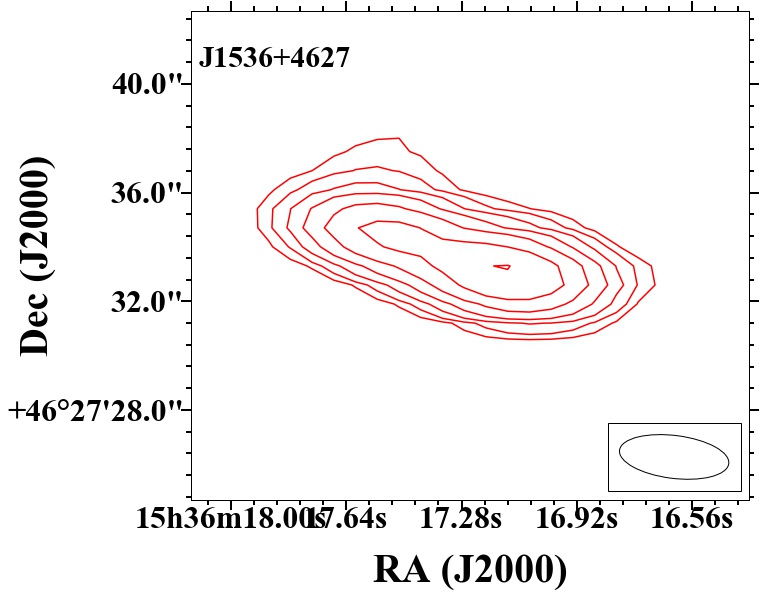}
\includegraphics[width=6cm, height=4.3cm]{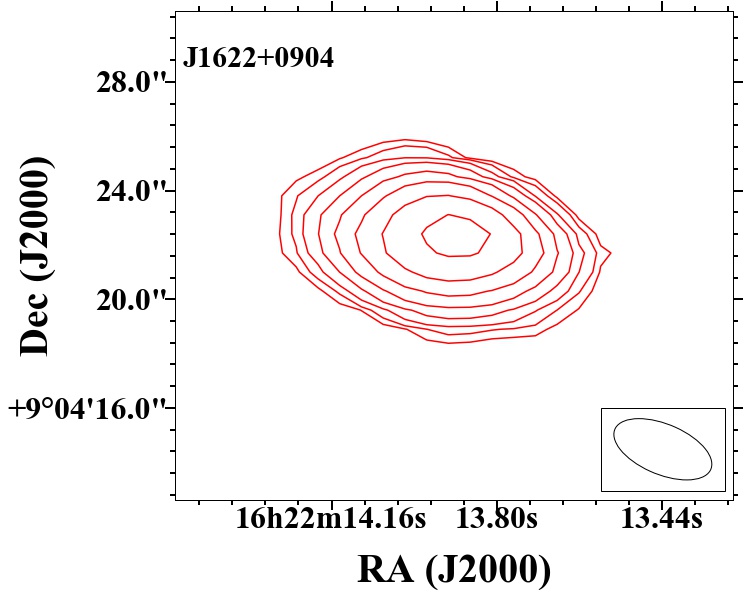}
\includegraphics[width=6cm, height=4.3cm]{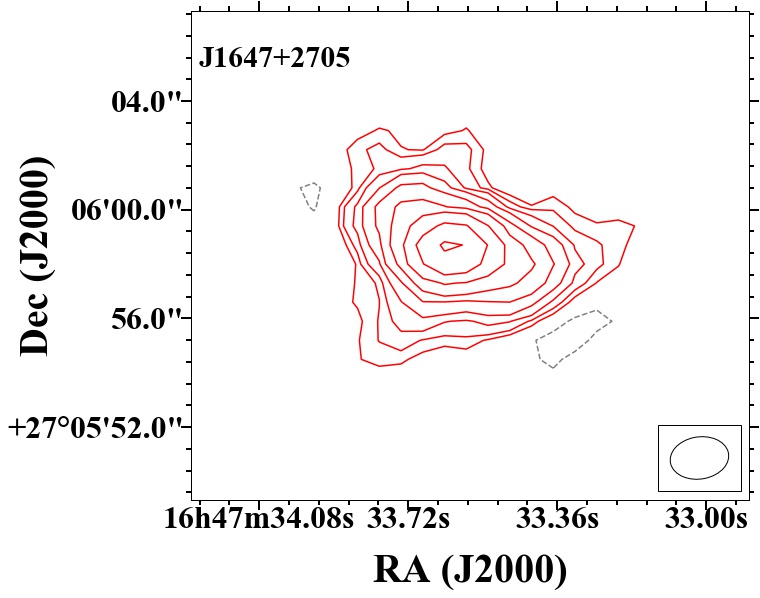}
\includegraphics[width=6cm, height=4.3cm]{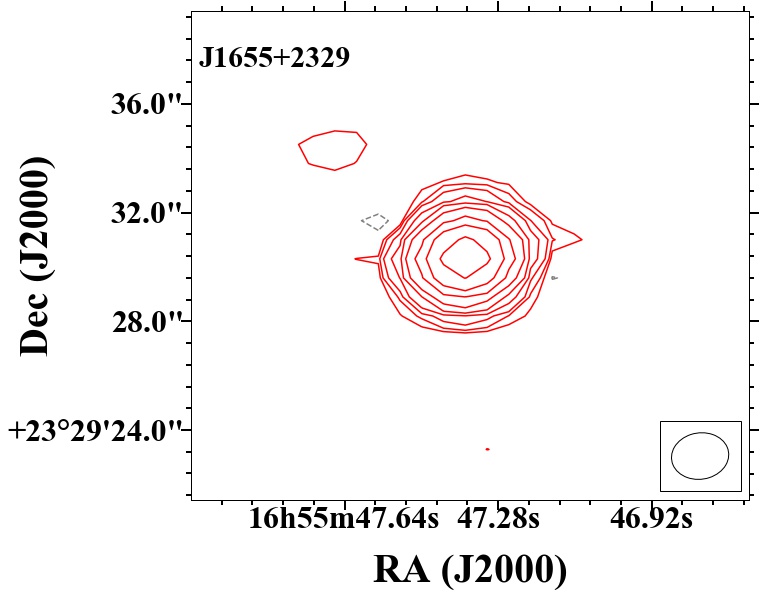}
\hspace{0.1cm}                                
\includegraphics[width=6cm, height=4.3cm]{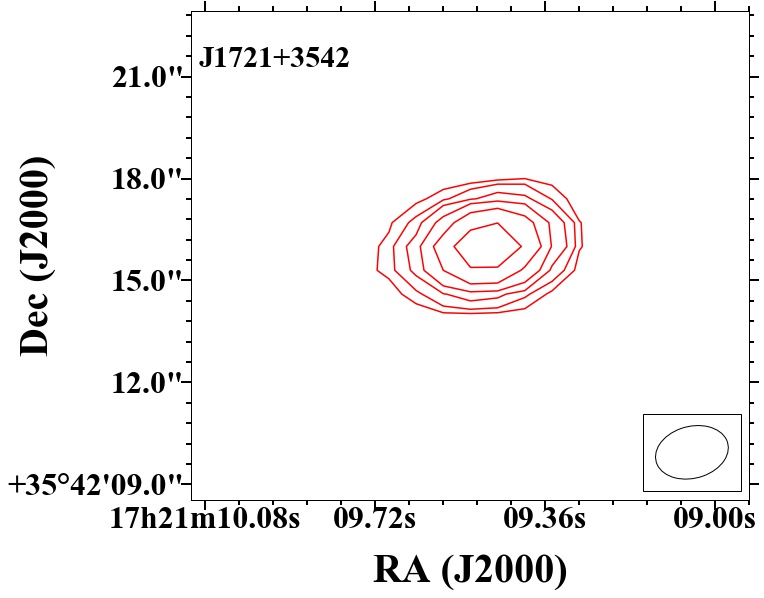}
\hspace{0.1cm}                               
\includegraphics[width=6cm, height=4.3cm]{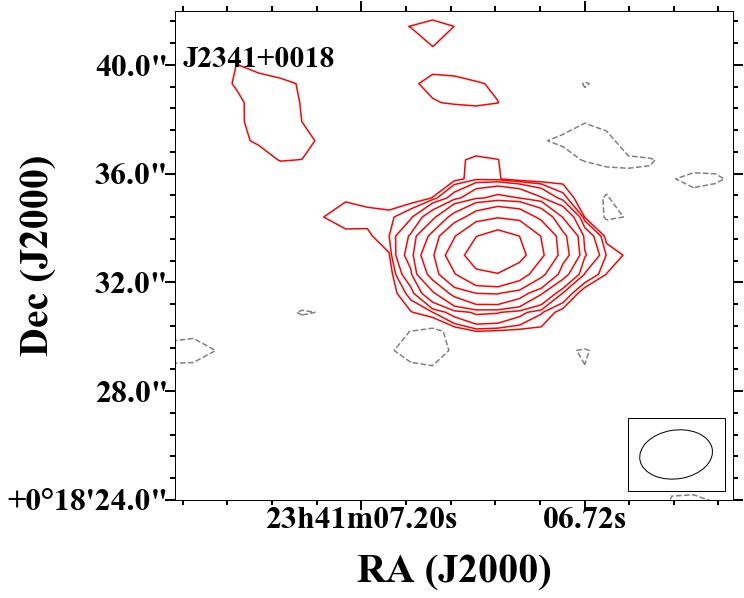}
   \caption{Continued.}
    \label{fig:continuum_maps}
\end{figure*}

\end{appendix}



\end{document}